\documentclass[pdflatex,sn-nature]{sn-jnl}

\usepackage{graphicx}%
\usepackage{multirow}%
\usepackage{amsmath,amssymb,amsfonts}%
\usepackage{amsthm}%
\usepackage{mathrsfs}%
\usepackage[title]{appendix}%
\usepackage{xcolor}%
\usepackage{textcomp}%
\usepackage{manyfoot}%
\usepackage{booktabs}%
\usepackage{algorithm}%
\usepackage{algorithmicx}%
\usepackage{algpseudocode}%
\usepackage{listings}%
\usepackage{rotating}
\usepackage{subcaption}
\usepackage{enumitem}
\usepackage{hyperref}
\usepackage{float}

\theoremstyle{thmstyleone}%
\theoremstyle{thmstyletwo}%
\theoremstyle{thmstylethree}%
\begin{document}

\title[Article Title]{Why ML-based cough models do not generalize: a systematic cross-dataset evaluation for tuberculosis screening}

\author*[1]{\fnm{Wensi} \sur{Zhang}}
\email{wensi.zhang@epfl.ch}

\author[2,3]{\fnm{Tomas} \sur{Teijeiro}}

\author[1]{\fnm{Jérôme} \sur{Thevenot}}

\author[1]{\fnm{David} \sur{Atienza}}

\affil[1]{\orgdiv{Embedded Systems Laboratory}, \orgname{École Polytechnique Fédérale de Lausanne}, \orgaddress{\state{Lausanne}, \country{Switzerland}}}

\affil[2]{\orgdiv{Basque Center for Applied Mathematics (BCAM)}, \orgaddress{\state{Bilbao}, \country{Spain}}}

\affil[3]{\orgdiv{University of the Basque Country (EHU)}, \orgaddress{\state{Leioa}, \country{Spain}}}


\abstract{
Cough acoustics are promising for non-invasive tuberculosis (TB) screening, yet whether machine learning (ML) models capture disease-related acoustics or artifacts of data collection remains unresolved. We evaluated the cross-dataset generalizability of classical ML and deep learning (DL) cough-based TB classifiers across three independent datasets. Despite moderate within-dataset performance (ROC-AUC up to $0.76 \pm 0.06$), both pipelines fail to generalize, with external performance frequently below 0.6, indicating a possible limitation of the data. We further observed that audio representations are organized by recording device and dataset rather than TB status, predicted TB probability tracks country-level prevalence in CODA, and device mismatch degrades transfer while device-diverse training improves it. Additionally, a clinical-variable baseline generalizes more consistently (ROC-AUC $0.65-0.71$), indicating acquisition-specific variability in acoustics is a stronger driver of poor generalizability. High within-dataset performance is not enough. External validation is essential before cough-based TB models are clinically ready.
}

\keywords{Tuberculosis Screening, Cough Monitoring, Acoustic Epidemiology, Machine Learning, Domain Generalization, Model Generalization}

\maketitle

\section*{Introduction}\label{sec1:intro}

Acoustic epidemiology is an emerging field that investigates how sound can provide clinically and epidemiologically relevant information. In respiratory health, growing interest has focused on the automated detection and analysis of cough for diagnosis, monitoring, and surveillance. The COVID-19 pandemic accelerated research in this area, leading to the development of machine learning (ML) models based on cough, speech, and breathing sounds, as well as large public audio collections such as COUGHVID~\cite{orlandic_coughvid_2021, despotovic_detection_2021, gabaldon-figueira_acoustic_2022}. However, robust validation has remained difficult. Public datasets are still limited, and large-scale crowd-sourced or loosely controlled collections introduce unreliable labels, variable recording environments, heterogeneous devices, and uncontrolled comorbidities~\cite{orlandic_coughvid_2021, ghrabli_challenges_2022, orlandic_semi-supervised_2023}. These factors create substantial domain shift and make it hard to determine whether a model has learned disease-related acoustics or artifacts of how the data were collected. This concern is not unique to cough. Across medical AI, models frequently achieve strong internal performance by exploiting spurious, acquisition-related ``shortcuts'' rather than disease signal~\cite{degrave_ai_2021, ong_ly_shortcut_2024, brown_detecting_2023, kechris_acoustical_2024}. 

Tuberculosis (TB) remains one of the world's leading infectious causes of death. In 2024, an estimated 10.7~million people developed TB and 1.23~million died~\cite{who_tuberculosis_nodate, who_global_nodate}. Confirmatory diagnosis typically requires sputum-based molecular or microbiological assays, often combined with chest imaging. These tools are difficult to deploy consistently in the low-resource settings that often carry the highest burden~\cite{zimmer_making_2022}. This has motivated strong interest in simple, scalable triage tools~\cite{who_target_nodate}.  
In this context, cough is of particular interest because it is both a common symptom of pulmonary TB and a main mechanism of transmission. A low-cost embedded device using cough could support screening, surveillance, and longitudinal monitoring~\cite{zimmer_making_2022}. 

Early TB-cough studies reported strong within-dataset results with area under the receiver operating characteristic curve (ROC-AUC) values around 0.95 for handcrafted features with classical classifiers~\cite{pahar_automatic_2021, botha_detection_2018}. Later studies explored more complex modeling strategies, including recurrent neural networks and attention-based architectures, with several reporting accuracies greater than 85\% or ROC-AUC values greater than 0.85~\cite{frost_tb_2022, yellapu_development_2023, rajasekar_detection_2024, xu_feature_2024}. However, evaluation was done almost always on small and private data under within-dataset validation. The release of large, structured TB-cough resources, such as the CODA dataset~\cite{huddart_dataset_2024}, the TBscreen dataset~\cite{sharma_tbscreen_2024}, and the Zambia CIDRZ dataset~\cite{baur_hear_2024}, now makes it possible to answer the question: does a model trained on one dataset retain performance on an independent dataset collected elsewhere?

Evidence that generalization of TB-cough models is challenging is already emerging. When the CODA TB DREAM Challenge models were validated on an independent Peruvian cohort collected under a closely matched protocol, ROC-AUC fell from 0.689--0.743 internally to 0.480--0.615 externally~\cite{zimmer_external_2026}. However, the cause of this decline was not further investigated. More broadly, cross-dataset evaluation in this field remains rare and methodologically inconsistent~\cite{pavel_tuberculosis_2025, kafentzis_tuberculosis_2026}. Known pitfalls, for example within-dataset-only validation, information leakage without subject-level splitting, and disease status confounded with site, device, or recruitment, are acute~\cite{xu_feature_2024, huddart_dataset_2024}.

\begin{figure}[t]
    \centering
    \includegraphics[width=\linewidth]{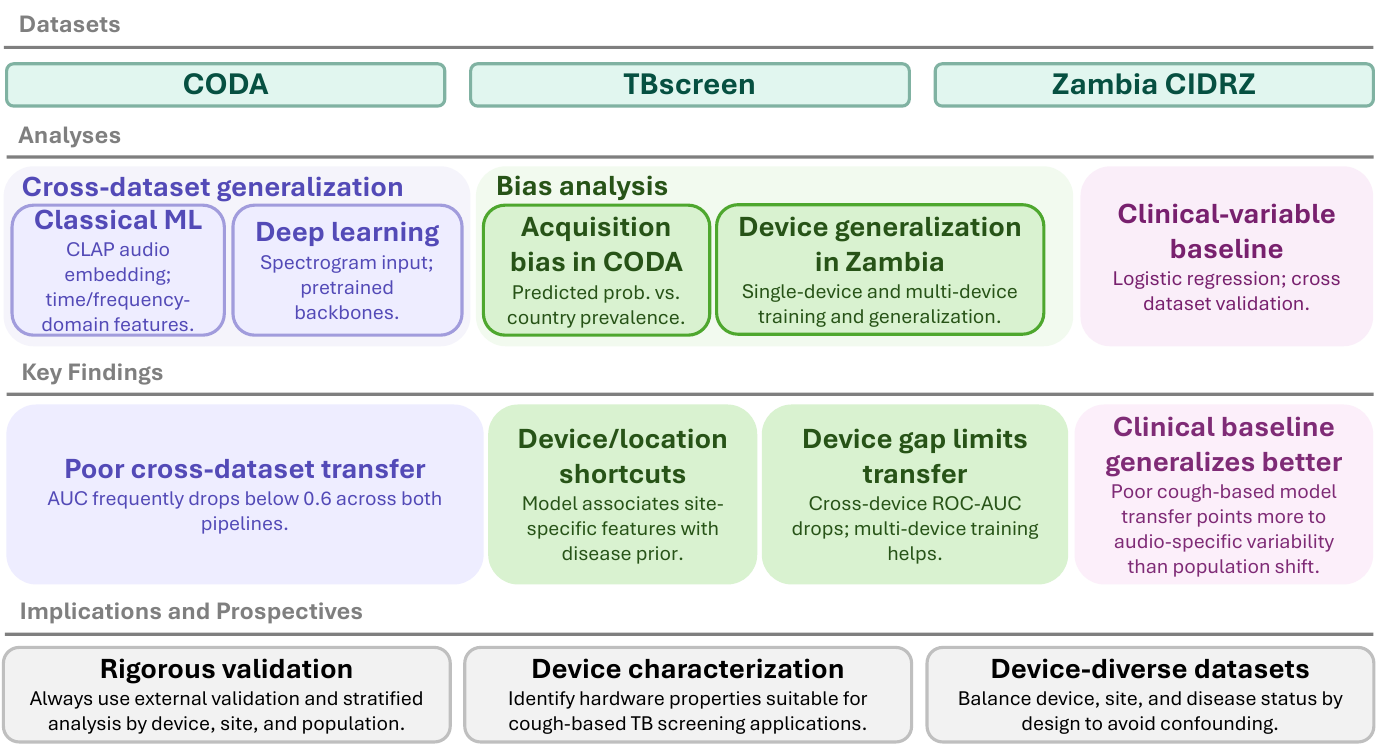}
    \caption{\textbf{Overview of the study design.} The study evaluates cross-dataset generalization of cough-based TB classifiers across three public datasets using classical ML and DL pipelines. Four major analyses are conducted: cross-dataset generalization, acquisition-related bias in CODA, device generalization in Zambia, and a clinical-variable baseline. Key findings and implications for future study design are also presented.}
    \label{fig:study_overview}
\end{figure}

Here we present a systematic cross-dataset evaluation of cough-based TB classifiers across the CODA, TBscreen, and Zambia datasets (Fig.~\ref{fig:study_overview}) using both a classical ML and a deep learning (DL) pipeline.
We complement this with an analysis of location- and device-related bias in the CODA dataset, a device-generalization study in the Zambia dataset, and a clinical-variable baseline. Our analysis yields four main messages. 
\vspace{-0.3cm}
\begin{enumerate}
    \item Moderate within-dataset performance does not transfer: cross-dataset ROC-AUC frequently drops below 0.6 for both pipelines, which reflects difference in data domains rather than the modeling approach.
    \item Models exploit acquisition shortcuts: audio representations are organized more strongly by device and dataset than by TB status and predicted TB probability tracks country-level prevalence in CODA.
    \item Recording-device mismatch directly degrades transfer, whereas device-diverse training improves robustness to unseen hardware.
    \item A simple clinical-variable model generalizes more consistently, pointing to audio-specific acquisition variability as a primary driver of poor transfer rather than population differences alone. 
\end{enumerate}
\vspace{-0.3cm}
Together, these results show that strong within-dataset performance should not be interpreted as evidence of clinical readiness.

\section*{Results}

\subsection*{Within-dataset performance does not transfer across datasets}

\begin{figure}[t]
    \centering
    \begin{subfigure}[t]{0.49\textwidth}
        \centering
        \includegraphics[width=\linewidth]{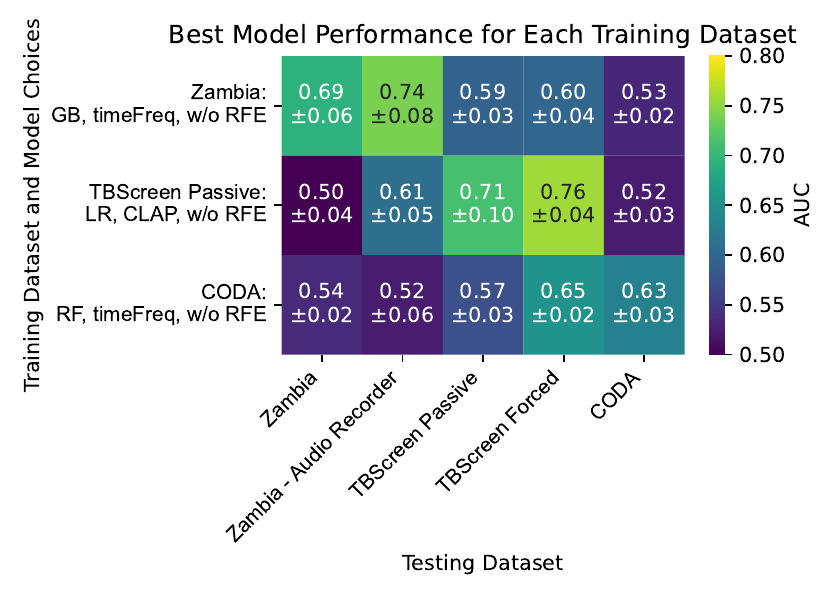}
        \caption{Classical ML pipeline}
        \label{fig:best_models_classical}
    \end{subfigure}
    \hfill
    \begin{subfigure}[t]{0.49\textwidth}
        \centering
        \includegraphics[width=\linewidth]{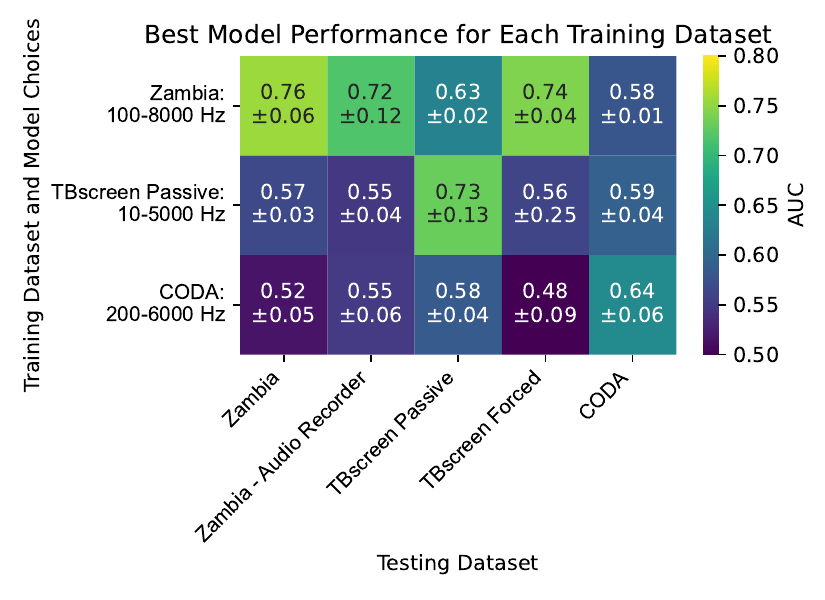}
        \caption{DL pipeline}
        \label{fig:best_models_dl}
    \end{subfigure}
    \vspace{0.8em}
    \begin{subfigure}[t]{0.5\textwidth}
        \centering
        \includegraphics[width=0.84\linewidth]{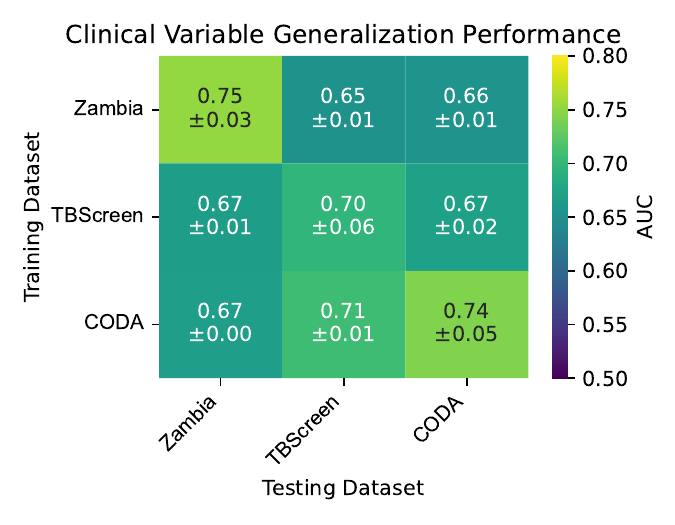}
        \caption{Clinical-variable baseline}
        \label{fig:clinical_variable_performance}
    \end{subfigure}
    \caption{
    \textbf{Cross-dataset performance of the acoustic pipelines and the clinical-variable baseline.}
    For each source dataset, the model configuration was selected using within-dataset validation and then evaluated on all target datasets. Values show subject-level ROC-AUC, with variability reported across outer folds.
    Across both acoustic pipelines, within-dataset performance did not translate into consistent external generalization, and the selected configuration differed by source dataset. In contrast, the clinical-variable baseline showed more stable external generalization across datasets.
    }
    \label{fig:best_models}
\end{figure}

We trained each pipeline on a single source dataset and evaluated it on the held-out datasets. Within the source dataset, we used nested cross-validation (CV). For classical ML pipelines, the inner folds were used for hyperparameter tuning and feature selection, and outer-fold performance guided the choice of preprocessing, feature set, and classifier. For DL pipelines, the inner folds were used for early stopping during training, and outer-fold performance guided the choice of hyperparameters, preprocessing, and input features. Training data were resampled to balance subgroups and limit model bias. The selected configuration was then evaluated, without further tuning, on each unseen target dataset. Performance is reported as the mean and standard deviation across the outer CV folds (Fig.~\ref{fig:best_models}), together with the 95\% confidence interval (CI) estimated by participant-level bootstrap (Extended Data Fig.~\ref{fig:best_models_ci}).

For the classical ML pipeline, within-dataset performance was moderate. The best model reached a subject-level internal ROC-AUC of $0.69 \pm 0.06$ (95\% CI $0.61-0.76$) on Zambia and $0.71 \pm 0.10$ (95\% CI $0.62-0.80$) on TBscreen passive cough, while CODA was lower at $0.63 \pm 0.03$ (95\% CI $0.58-0.68$). External performance was consistently weaker. Models trained on Zambia or TBscreen performed well on their own data but did not transfer reliably to the other datasets. 
The best-performing configuration also differed by source dataset, and no single classifier, feature family, or feature-selection setting dominated.
Performance varied far more with the choice of training and testing dataset than with any modeling choice (Extended Data Fig.~\ref{fig:classical_ml_grand_performance}), which indicates that the limiting factor is the mismatch between datasets rather than the pipeline itself.

The DL pipeline followed the same pattern at a slightly higher performance level. A VGGish backbone~\cite{hershey_cnn_2017} with mel-spectrogram input performed best and was used throughout. The Zambia-trained model reached the strongest within-dataset performance, with a ROC-AUC of $0.76 \pm 0.06$ (95\% CI $0.70-0.80$), and transferred moderately to the Zambia audio-recorder subset ($0.72 \pm 0.12$, 95\% CI $0.65-0.78$) and to TBscreen forced cough ($0.74 \pm 0.04$, 95\% CI $0.67-0.80$), but it fell to $0.63 \pm 0.02$ (95\% CI $0.60-0.67$) on TBscreen passive cough and $0.58 \pm 0.01$ (95\% CI $0.56-0.60$) on CODA. Models trained on TBscreen passive cough (within-dataset $0.73 \pm 0.13$, 95\% CI $0.66-0.80$) or on CODA (within-dataset $0.64 \pm 0.06$, 95\% CI $0.60-0.68$) did not generalize, with external ROC-AUC falling below $0.6$. 
We further examined whether these within-dataset results were uniform across acquisition subgroups, stratifying by recording device and location for each source dataset, to complete the analysis (Extended Data Fig.~\ref{fig:Zambia_spec}~-~Fig.~\ref{fig:coda_spec_device}).

The first key finding is that models do not generalize well for both classical ML and DL approaches. This is likely a result of domain shift between datasets. This hypothesis was explored next by representation analysis.

\subsection*{Acoustic representations are organized by acquisition factors rather than TB status}

To understand why the models generalized poorly, we examined the structure of the input feature spaces directly. We quantified differences between subgroups using maximum mean discrepancy (MMD) and visualized the resulting pairwise distance structure with multidimensional scaling (MDS). We also projected individual samples with t-distributed stochastic neighbor embedding (t-SNE) to inspect how the feature spaces were organized. Subgroups were defined by recording device, location, or TB label. 

\begin{figure}[t]
    \centering
    \begin{subfigure}[t]{0.32\textwidth}
        \centering
        \includegraphics[width=\linewidth]{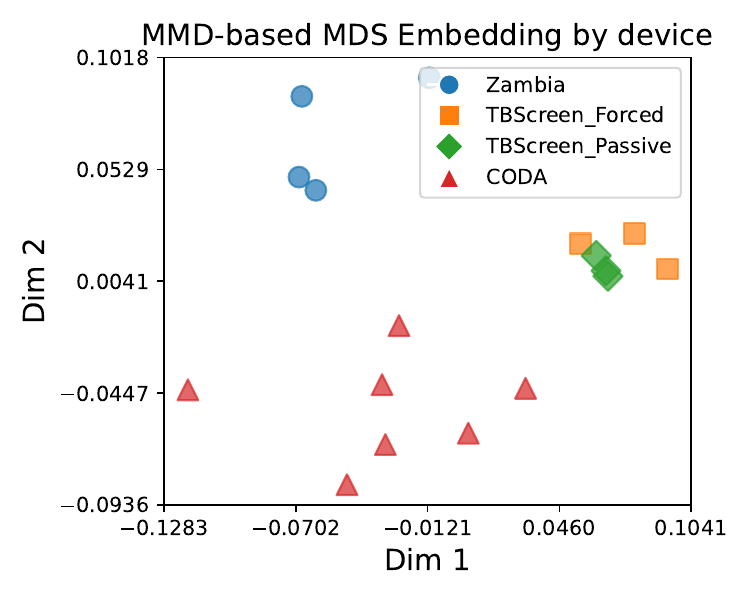}
        \caption{Device}
        \label{fig:mmd_clap_device}
    \end{subfigure}
    \hfill
    \begin{subfigure}[t]{0.32\textwidth}
        \centering
        \includegraphics[width=\linewidth]{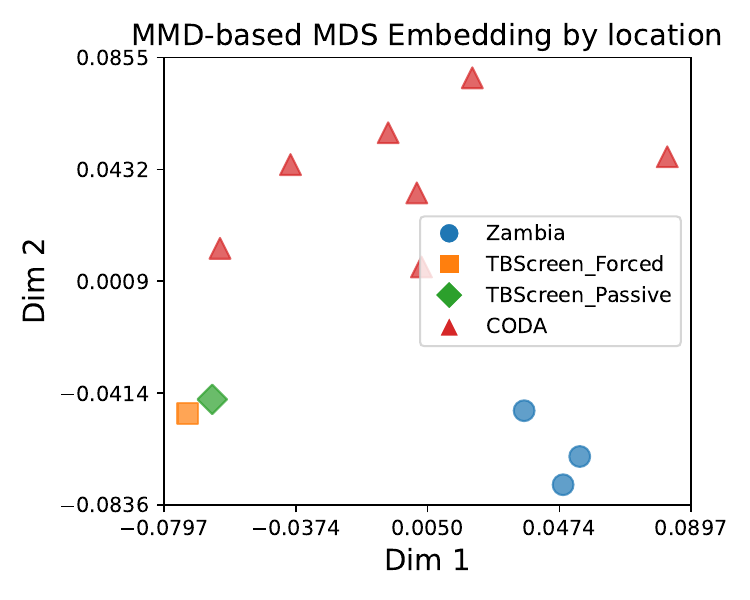}
        \caption{Location}
        \label{fig:mmd_clap_location}
    \end{subfigure}
    \hfill
    \begin{subfigure}[t]{0.32\textwidth}
        \centering
        \includegraphics[width=\linewidth]{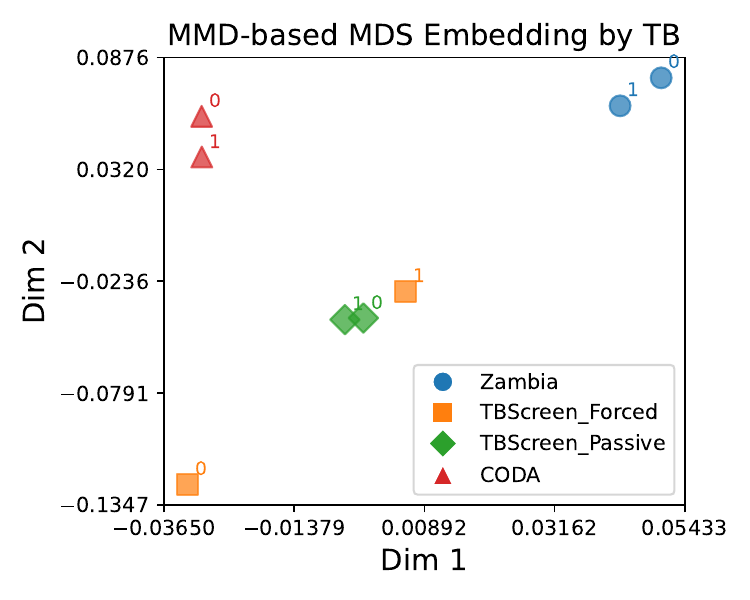}
        \caption{TB label}
        \label{fig:mmd_clap_tb}
    \end{subfigure}

    \caption{
    \textbf{MMD-based MDS visualization of subgroup structure in CLAP audio embeddings.}
    Pairwise subgroup distances were computed using MMD and visualized using MDS. The axes do not have a physical meaning. They represent latent dimensions that preserve pairwise similarities, with points closer together indicating more similar data distributions. The different scatter points are colored by the dataset source, and they represent subgroups defined by recording device, location, or TB label. Subgroups defined by TB label were labeled in \textit{(c)} to indicate positive (1) or negative (0) status. In CODA, location and device are overlapping factors because country source is associated with recording-device choice. Ideally, device/location should not cause much separation and subgroups by device/location of different dataset sources should overlap, whereas subgroups by TB status should cluster by their TB labels. However, it is the opposite in reality.
    }
    \label{fig:mmd_clap_subgroups}
\end{figure}

The dominant structure in the feature spaces was related to acquisition rather than to disease. In the MMD-MDS visualizations for CLAP embeddings, sample subgroups (the scatter points in Fig.~\ref{fig:mmd_clap_subgroups}) from different dataset sources barely overlap, but rather cluster by themselves. Subgroups defined by device (Fig.~\ref{fig:mmd_clap_device}) and location (Fig.~\ref{fig:mmd_clap_location}) span a large area, even when they are from the same dataset. Meanwhile, TB+ and TB- subgroups of the same dataset (Fig.~\ref{fig:mmd_clap_tb}) are often similar. These indicate that the clearest separation of the data representation corresponds to dataset source and recording device, not to TB status. Similar effects can be observed for handcrafted time-frequency features (Supplementary Fig.~B2).

The t-SNE projections show the same pattern (Supplementary Fig.~B3--B6). Samples form broad clusters correspond to dataset source, and within each dataset there is additional structure related to recording device, whereas separation by TB label is comparatively weak. Together, these visualizations indicate that the acoustic features carry strong acquisition-related information.

This provides a mechanism for the generalization failure reported above and supports our second key result. When acquisition factors are this prominent in the feature space, and when they are correlated with TB status within a dataset, a classifier can fit decision boundaries that depend on dataset- or device-specific structure rather than on disease-related cough characteristics that are stable across settings.

\subsection*{In CODA, predicted TB probability tracks country-level prevalence}

\begin{figure}[t]
    \centering
    
    \begin{subfigure}[t]{0.49\textwidth}
        \centering
        \includegraphics[width=1.05\linewidth]{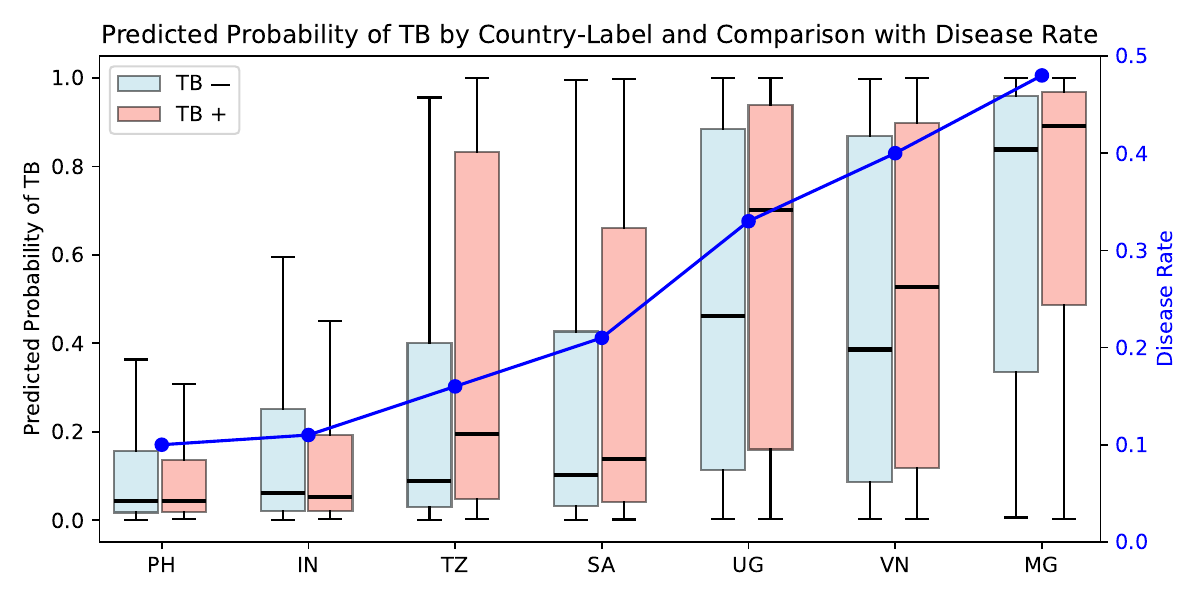}
        \caption{Challenge-replication setting}
        \label{fig:coda_pred_prob_replication}
    \end{subfigure}
    \hfill
    \begin{subfigure}[t]{0.49\textwidth}
        \centering
        \includegraphics[width=1.05\linewidth]{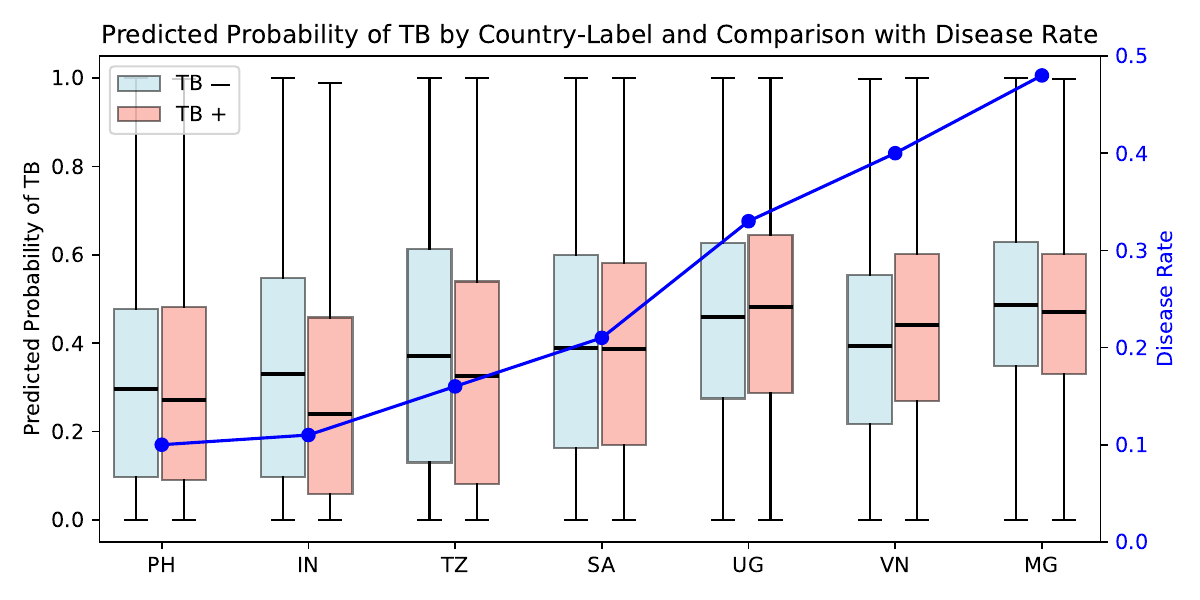}
        \caption{Conservative setting}
        \label{fig:coda_pred_prob_conservative}
    \end{subfigure}
    
    \caption{
    \textbf{Predicted TB probabilities by country source and TB label in CODA.} 
    The blue line shows the country-level TB+ rate.
    \textbf{(a)} Under the challenge-replication setting, predicted TB probabilities increased strongly with country-level TB prevalence for both TB+ and TB- samples.
    \textbf{(b)} Under the conservative setting, predicted probabilities were less extreme and the association with country-level prevalence was reduced, although not fully removed.
    }
    \label{fig:coda_pred_prob_country_label}
\end{figure}

The representation analysis showed that acquisition structure dominates the feature space. We next investigated whether this structure has a direct, measurable effect on model output, using the multi-country CODA dataset, where each country is associated with a particular choice of recording device(s) and a particular TB+ rate. We examined whether predicted TB probabilities were associated with country-level TB prevalence under two modeling settings. The first setting closely replicated the best-performing CODA TB DREAM Challenge pipeline~\cite{jaganath_accelerating_2025}, using a ResNet-34 model with high-frequency spectrogram inputs and resampling at the TB-label level only. The second was a more conservative setting designed to reduce acquisition-related shortcuts, using a VGGish-based model, a reduced upper frequency limit of 8,000~Hz, cough segmentation before feature extraction, and resampling jointly by TB label and country source. Because several design choices differ between the two settings, they are not intended as a controlled comparison of individual factors. Rather, they represent two configurations selected with different priorities, the first maximizing predictive performance and the second balancing performance against acquisition-related bias.

Under the challenge-replication setting, the model output depends strongly on country source rather than on disease state. Countries with higher TB+ rates receive higher predicted TB probabilities, and this holds for both TB+ and TB- samples (Fig.~\ref{fig:coda_pred_prob_replication}). Regressing the mean predicted probability per country against the country-level disease rate gives a slope of 1.20 with $R^2 = 0.95$ for TB- cases and a slope of 1.30 with $R^2 = 0.89$ for TB+ cases, so country-level prevalence alone accounts for most of the variance in mean model output within each group. The model behaves as if it is estimating population-level prevalence from acquisition cues rather than detecting disease in individual coughs. This is problematic in practice: under a single fixed decision threshold, the model would predict nearly all subjects from a high-prevalence country as TB+ and nearly all subjects from a low-prevalence country as TB-, regardless of the disease-related acoustic content of the individual cough.

Under the conservative setting, this association is attenuated but not fully removed. 
TB- cases give a regression slope of 0.33 and an $R^2$ of 0.75, and TB+ cases give a regression slope of 0.33 and an $R^2$ of 0.84. The predicted probabilities are less extreme and span more similar ranges for both TB+ and TB- groups across countries (Fig.~\ref{fig:coda_pred_prob_conservative}). 
Some dependence on prevalence is not in itself a defect, since prior disease rates carry information that can aid classification in deployment. The problem is when such priors are absorbed through acquisition shortcuts rather than from disease-relevant acoustics, they become a source of miscalibration when a model is moved to a population with different prevalence.

These findings support our second key result and motivate the subgroup-balanced resampling strategy used in this study. Label-only resampling is insufficient in a heterogeneous dataset such as CODA: a model can learn to associate acquisition factors with country-level prevalence. This shows why generic performance evaluation is not sufficient. Post-hoc analysis of model outputs across sites, devices, and other factors is essential to expose residual bias that an overall evaluation can hide.

\subsection*{Recording-device mismatch degrades transfer, while device-diverse training improves robustness}

\begin{figure}[t]
    \centering
    \begin{subfigure}[t]{0.98\textwidth}
        \centering
        \includegraphics[width=\linewidth]{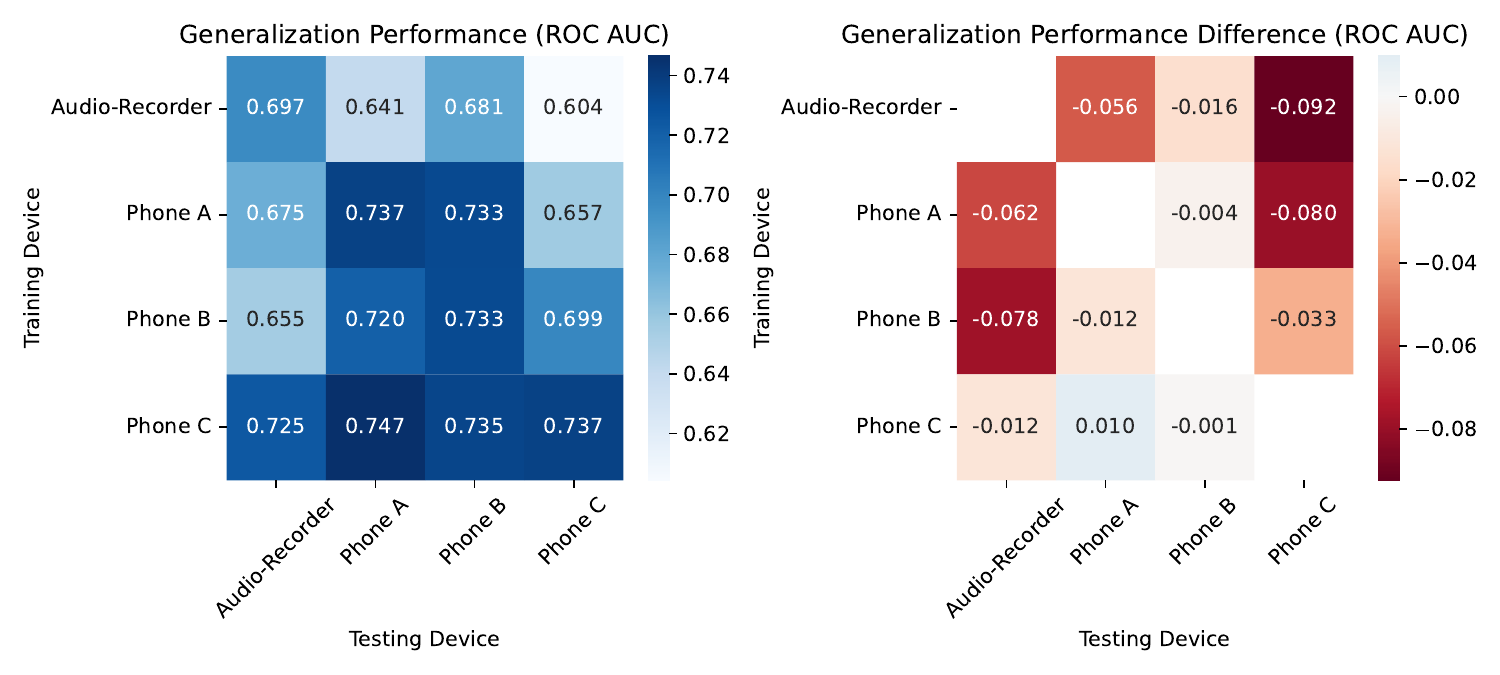}
        \caption{Single-device training.}
        \label{fig:zambia_device_a}
    \end{subfigure}
    \vspace{0.8em}
    \begin{subfigure}[t]{0.98\textwidth}
        \centering
        \includegraphics[width=0.49\linewidth]{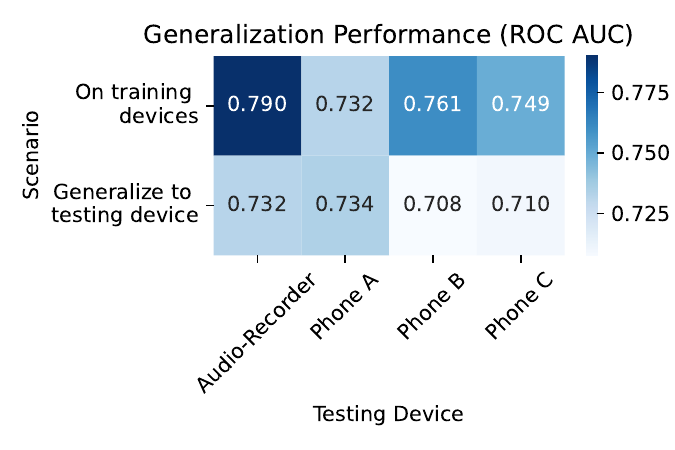}
        \hfill
        \includegraphics[width=0.49\linewidth]{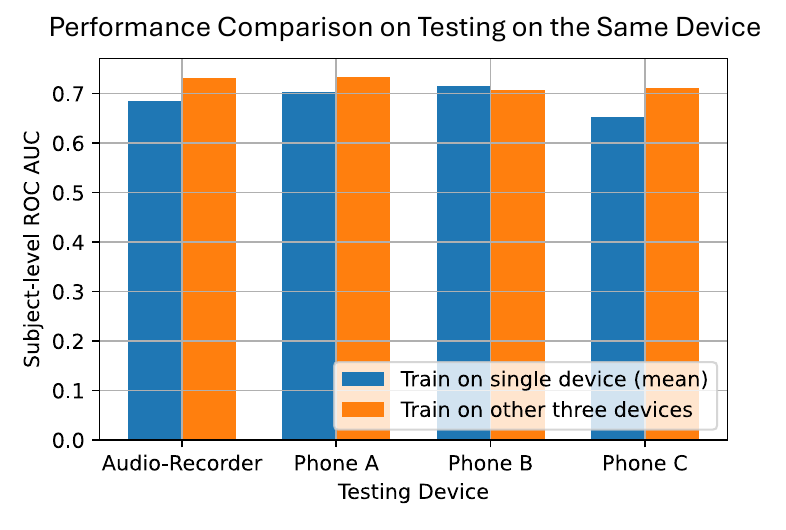}
        \caption{Multi-device training.}
        \label{fig:zambia_device_b}
    \end{subfigure}
    \caption{
    \textbf{Recording-device effects on generalization in the Zambia dataset.}
    Values are subject-level ROC-AUC.
    \textbf{(a)} Single-device training. The left heatmap reports ROC-AUC for each training-device and testing-device pair, with diagonal entries showing within-device evaluation and off-diagonal entries showing cross-device evaluation; the right heatmap shows the change in ROC-AUC of each cross-device evaluation relative to the corresponding within-device value. Cross-device evaluation generally reduced performance.
    \textbf{(b)} Multi-device training. The left heatmap reports ROC-AUC for models trained on three devices and evaluated on either the training devices or the held-out device. The right plot compares held-out-device performance between single-device training and training on the other three devices, where the single-device value is the average over the three possible single-device choices. Training across multiple devices generally improved performance on the unseen device.
    }
    \label{fig:zambia_device}
\end{figure}

The CODA results show that classifiers can tie disease labels to acquisition structure, but in CODA, country, device, recruitment setting, clinical population, and prevalence are entangled. We therefore used the Zambia dataset, in which recordings from the same participants were often captured on several devices simultaneously, to test device effects more directly. 
The devices include one audio recorder and three phone models. Two experiments were conducted. In the single-device experiment, a model was trained with samples from one device and applied to all four devices. In the multi-device experiment, a model was trained on samples from three devices and applied to either unseen samples from the training devices or samples from the held-out device.

Recording-device mismatch reduces performance. In the single-device experiment, cross-device evaluation is generally worse than within-device evaluation, shown by the negative off-diagonal differences in subject-level ROC-AUC (Fig.~\ref{fig:zambia_device_a}). The extent of the performance drop varies by device.  
Among the devices, Phone C is the strongest and most consistent training device. 

Exposure to multiple devices during training improves robustness to unseen hardware. The device-mismatch effect remains in the multi-device experiment, but training on three devices generally improves performance on the held-out device relative to single-device training (Fig.~\ref{fig:zambia_device_b}). For example, with the audio recorder as the held-out device, training on Phone~A, Phone~B, or Phone~C individually gives ROC-AUC values of 0.675, 0.655, and 0.725, while training on all three together raises it to 0.732. This gain is not simply a matter of more data: Phone~A, Phone~B, and Phone~C cover 632, 628, and 642 unique subjects respectively, and the three combined cover 650, so the added volume is device diversity rather than subject diversity. The improvement therefore reflects exposure to multiple recording devices, which helps the model learn representations that are less specific to any one device.
To further rule out a sample-size effect, we repeated the experiment with the training data from the three remaining devices resampled to match the number of samples in the single-device setting. The multi-device model still transferred better to the held-out device, indicating that the gain stems from device diversity rather than from additional training samples (Extended Data Fig.~\ref{fig:device_multi_matched}).

These experiments support our third key result. Recording-device mismatch contributes directly to the limited generalization of cough-based TB classifiers, since a model trained on one device may not transfer to another even within the same study population and sites, and device-diverse training mitigates this effect. 

\subsection*{A clinical-variable baseline generalizes more consistently than the acoustic models}

To test whether the generalization failure is specific to audio or reflects population differences more broadly, we evaluated a non-acoustic baseline built from routinely collected clinical variables. Using only features available in comparable form across all three datasets, namely age, sex, smoking status, HIV status, previous TB history, cough duration, fever, weight loss, and night sweats, we trained a logistic-regression model on one dataset and evaluated it on the others, under the same external-validation framework.

The clinical-variable model transfers more consistently than the acoustic pipelines (Fig.~\ref{fig:clinical_variable_performance}). For instance, trained on Zambia, the model reached a within-dataset ROC-AUC of $0.75 \pm 0.03$ (95\% CI $0.71-0.78$) and external $0.65 \pm 0.01$ (95\% CI $0.60-0.71$) on TBscreen and $0.66 \pm 0.01$ (95\% CI $0.64-0.67$) on CODA. External performance is not uniformly high, but it stays consistently above chance and falls within a narrower range than the acoustic-model results. The contributions of individual clinical variables, assessed by model coefficients and permutation importance, are reported in Extended Data Fig.~\ref{fig:clinical_feature_importance}. Some symptoms such as weight loss were stable across datasets, while others were dataset-dependent, indicating that clinical variables are not entirely free from dataset shift either.

This contrast supports our fourth key result and sharpens the interpretation of the whole analysis. Structured clinical information, which also differs across these cohorts in population, recruitment, and prevalence, nonetheless transfers more stably than the cough acoustics. Audio-specific acquisition variability is likely the primary driver of the poor transfer of cough-based models, rather than population shift alone.

\section*{Discussion}

This study examined whether cough-based ML models for TB classification generalize across independently collected datasets. Across both the classical ML and DL pipelines, within-dataset performance does not translate into external performance. We observe moderate internal ROC-AUC, up to 0.755 for DL and 0.711 for classical ML. However, in cross-dataset evaluation, performance frequently drops below 0.6. This indicates that the limitation is a property of the data and its acquisition rather than of any particular modeling approach. This is not specific to our pipelines. An independent external validation of the CODA TB DREAM Challenge models on a Peruvian cohort reported external ROC-AUC of 0.480--0.615, even though the models were achieving ROC-AUC of 0.689--0.743 internally and the Peru data were collected under a closely matched acquisition protocol~\cite{zimmer_external_2026}. The study mainly attributed this challenge to population shift. 
Our results suggest that population shift likely accounts for only part of it, and that acquisition-related factors also play a substantial role. Here we isolate one such factor, recording-device variation, and show that it alone can markedly degrade transfer. Other confounding factors likely contribute as well, but they cannot be disentangled with the datasets currently available.

A mechanism for this failure is visible directly in the representations. In the datasets analyzed, differences in recording device, site, recruitment setting, cough protocol, and population are often larger and more stable than the acoustic differences associated with TB status. There can be two effects. First, the variation introduced by acquisition factors can overwhelm the variation associated with TB status in the feature space. A model then fits a decision rule that is locally valid within the source dataset, where acquisition conditions are approximately fixed, but may not remain valid once the domain shifts substantially. Second, when acquisition factors are correlated with TB status, the model can learn these factors directly as shortcuts for prediction. 

The CODA prevalence analysis makes this shortcut concrete. Under label-only balancing, predicted TB probability tracks country-level prevalence for both TB+ and TB- samples. Cough segmentation, a reduced frequency range, a simpler model, and joint balancing by label and acquisition-factor-defined subgroups attenuate this association. A model that relies on such shortcuts cannot transfer to settings where they no longer hold, which offers an explanation for why CODA-derived models struggled when validated externally on the Peru cohort~\cite{zimmer_external_2026}.

The Zambia experiments isolate one such factor. Recording-device mismatch directly degrades transfer. Training across multiple devices improves performance on an unseen device. However, as the CODA analysis shows, this strategy can be beneficial when device diversity is introduced within a balanced study design. If device type is confounded with location, disease prevalence, recruitment pathway or other factors, adding more devices may increase dataset size while also creating new shortcuts for the model.

These observations bound what algorithmic remedies can achieve. The established domain-generalization methods we evaluated (listed in Supplementary~C) gave no consistent external improvement, likely because such methods require sufficiently large and well-balanced source domains to separate stable task-relevant structure from domain-specific variation. In the available TB-cough datasets, domain, device, site, prevalence, recruitment, and participant characteristics are partially confounded, and the TB-related acoustic signal appears weak relative to device and protocol effects, so enforcing domain invariance may discard useful information without improving the clinically relevant boundary. Algorithmic domain generalization is therefore unlikely to compensate for limitations in dataset design.

The clinical-variable baseline provides an important point of reference. A simple logistic regression model using basic clinical variables generalizes more consistently across datasets than any acoustic model. This suggests that the poor transfer of cough-based models cannot be explained only by differences in population. Instead, audio-specific acquisition variability plays a major role. This motivates multimodal models in which clinical variables supply a stable baseline risk estimate while cough acoustics contribute complementary information, reducing the need for the audio model to learn the full decision boundary and its reliance on dataset-specific shortcuts. Additionally, clinical variables such as previous TB, HIV status, smoking, or symptom profile may help define subgroups with more homogeneous acoustics. Finally, cough frequency, bout structure, and temporal dynamics, beyond isolated single coughs, may also carry more transferable and clinically meaningful information.

We identify two main paths for future cough-based TB screening research. The first is a data-driven, smartphone-oriented strategy that collects much larger datasets across many devices, environments, countries, and participant groups, so that models can learn representations robust to realistic deployment variation. The second is a standardized-device strategy that collects recordings on a small set of devices with known and stable acoustic properties under a fixed protocol, reducing acquisition variability and making model behaviour easier to interpret. The Zambia device analysis provides some support for this idea: among the devices, Phone C produced the strongest within-device performance and transferred best to others. 
This is notable because it contradicts the simple expectation that a dedicated audio recorder, with a flatter frequency response, should provide the best training signal. Device suitability appears to depend not only on nominal recording quality but also on how a device captures cough-relevant structure and how similar its recordings are to those from other devices.
Future work should therefore examine device frequency response, automatic gain control, compression, microphone placement, and noise-processing characteristics to identify which hardware properties are most suitable for cough-based screening.

We also explored using health acoustics foundation models as feature extractors, for example OPERA~\cite{zhang_towards_2024}, but they did not outperform the pipelines reported here. This is consistent with Niizumi et al.~\cite{niizumi_towards_2025}, who found that models pre-trained on AudioSet, a large general-purpose audio corpus, were generally more effective for respiratory tasks than models pre-trained only on respiratory audio. Several factors may explain this. First, respiratory pre-training corpora remain small and less diverse than general audio corpora, which limits what self-supervised objectives can learn from them. Second, these corpora are typically assembled from a few studies, each with its own devices, recording settings, and protocols, so the learned representations may inherit the same acquisition-specific characteristics that drive the shortcuts identified in this work. Third, standard pre-training objectives are not designed to separate disease-related acoustic cues from recording conditions, so domain-specific pre-training alone does not guarantee robustness to acquisition shift. Together, these observations suggest that larger health-specific models are unlikely to resolve the generalization problem unless their pre-training data and objectives explicitly account for acquisition variability.

Causal analysis offers a complementary direction. For models based on demographic and clinical variables, a key question is which variables are direct causal parents or children of TB status and whether their relationships with TB remain invariant across settings. Identifying such stable causal relationships would provide a principled basis for screening models that stay robust when populations and health systems change. For cough acoustics, causal analysis could also help determine whether a given acoustic representation is a downstream effect of TB itself, or instead a descendant of the recording environment, the device, or demographic factors. This distinction would allow acquisition-driven features to be identified and excluded before deployment, rather than discovered after external validation fails. Such analyses will, however, require datasets in which these factors are measured and sufficiently varied.

This study has several limitations. For both acoustic pipelines, pipeline-level choices were selected using the mean outer-fold performance, so internal ROC-AUC values may be optimistic and should be read as upper bounds. However, external estimates are not affected. Moreover, although the datasets are broadly comparable, they differ in TB-labeling criteria, acquisition protocol, recording device, inclusion criteria, and population, and these residual differences cannot be fully separated from the generalization effects we report. A specific aspect is the difference in ethnicity and geographic origin. Variation in airway and vocal-tract anatomy, body size, respiratory comorbidity patterns, and culturally shaped coughing behaviour could all give rise to acoustic differences between, for example, Asian and African participants, independently of TB status. 
The public releases of datasets were sometimes partial, so some analyses used subsets that may not represent the parent cohorts. The available external cohorts are also few, and additional settings with different TB burden, infrastructure, or device ecosystems would be needed to test how broadly these patterns hold. Finally, our representation analysis demonstrated that the dataset and device structure dominated the acoustic feature spaces, but we did not investigate in depth which specific acoustic properties drove this separation, partly because consumer-smartphone microphone hardware and signal processing are undocumented and vary across conditions.

Shortcut learning is a well-recognized problem in ML and in medical AI in particular, but to our knowledge it has not been systematically examined for cough-based TB screening. This study provides such an analysis, showing how acquisition factors can enter cough models as shortcuts and identifying plausible sources of the resulting domain shift. The picture that emerges is sobering: reliable cough-based screening is considerably harder to achieve than internal validation scores suggest.
These findings do not, however, imply that cough acoustics are uninformative for TB screening. The fact that external validation often gives above-chance performance suggests that cough acoustics contain information on TB status. Rather, current dataset designs and validation practices make it difficult to determine whether a model has learned disease-relevant acoustic features, and strong internal performance is not equivalent to clinical readiness. 
For researchers designing new studies, this means treating device, site, recruitment pathway, and recording protocol as potential confounders from the outset, documenting them in detail, and balancing them across disease status wherever possible. Progress toward deployment will depend less on raising internal validation scores and more on prospective external validation, balanced acquisition designs, and evaluation under realistic deployment conditions.
More broadly, the validation and analysis pipeline proposed in this study is not limited to TB-cough, but can be extended to any ML-based acoustic classifier. It provides a practical framework for evaluating model generalizability and identifying the sources of performance degradation.

\clearpage
\section*{Methods}\label{sec3:methods}

This study first investigated whether cough-based TB classifiers generalize across independently collected datasets using a classical ML and a DL approach. We then explored further into the location-related or device-related bias in specific datasets. A clinical-variable model was also trained to establish a performance reference.

\subsection*{Training and evaluation design}

The pipelines were specifically designed to avoid data-leakage and overestimated performance measurements.
Firstly, all within-dataset models were trained and validated using nested CV, see Supplementary~D. The procedure used 5 outer folds and 4 inner folds, and was repeated twice with different random splits.
The performance of the model was assessed using ROC-AUC, and the results are reported as the mean and standard deviation across the 10 outer-fold estimates. 
95\% CI for ROC-AUC were also estimated using a participant-level bootstrap. Predictions were first aggregated to one score per participant by taking the median of the predicted scores of their individual coughs. Because each outer fold has its own trained model, and predicted scores may therefore not be comparable across models, resampling was performed within each fold rather than on pooled predictions. In each of 2,000 bootstrap replicates, TB-positive and TB-negative participants were resampled separately with replacement within each fold, preserving the fold’s TB prevalence, and the ROC-AUC was computed per fold and averaged across folds. The 2.5th and 97.5th percentiles of the resulting distribution of mean ROC-AUC were taken as the interval bounds, while the point estimate is the mean ROC-AUC across all folds.
For cross-dataset experiments, models of each iteration were evaluated directly on the held-out external dataset without cross-validation.

Secondly, all model-development experiments used subject-level data splitting. This was essential because each participant could contribute multiple cough recordings, and allowing coughs from the same participant to appear in both training and validation sets would introduce information leakage and lead to overly optimistic performance estimates. Therefore, all folds in the nested CV procedure were generated at the subject level rather than at the cough-event level. After model training, each cough event from a participant was assigned a predicted TB probability. Then these cough-level probabilities were summarized by their median across all coughs from the same participant to obtain a single subject-level prediction. Model performance was subsequently assessed using ROC-AUC, computed from these subject-level predictions and the corresponding subject-level TB labels.

Class imbalance was also considered during model development. Because TB+ and TB- participants were not equally represented in all datasets, class-balancing strategies need to be taken, which can be class weighting, oversampling of the minority class, or under-sampling of the majority class. As TB+ samples are often scarce, oversampling of minority class in the training data was applied.

However, class imbalance alone does not capture all sources of potential bias. In heterogeneous cough datasets, disease status may be correlated with non-disease factors such as recording site, country, device, or clinical population. In such cases, balancing only the TB+ and TB- classes may still allow a model to exploit dataset-specific or acquisition-specific cues that are indirectly associated with TB prevalence. A model trained with class-balanced CODA data was found to make predictions that traced location-specific TB prevalence (Fig.~\ref{fig:coda_pred_prob_replication}). As a result, a more cautious resampling strategy was taken. Rather than balancing only the TB+ and TB- classes, we balanced the training data within acquisition-defined subgroups. For each dataset, subgroups were defined by the combination of recording location, device, and TB status whenever the factors were available. Upsampling was applied to balance the number of cough samples across these subgroups. By enforcing a more balanced representation of acquisition-defined subgroups during training, we aimed to limit the extent to which models could rely on subgroup prevalence or device-specific acoustic characteristics as shortcuts for TB classification.

\subsection*{Datasets}
We considered three publicly available TB cough datasets that are broadly comparable in their clinical objective and overall study design: the CODA dataset~\cite{huddart_dataset_2024}, the TBscreen dataset~\cite{sharma_tbscreen_2024}, and the Zambia CIDRZ dataset~\cite{baur_hear_2024}. This makes them well suited for testing our central hypothesis that cough acoustics contain disease-relevant information that should support the development of models with meaningful cross-dataset generalization. A comparison of the three datasets, including their study populations, recording setups, and diagnostic reference standards, is provided in Supplementary~E.

The CODA dataset is a collection assembled for the CODA TB DREAM Challenge from seven countries across Asia and Africa~\cite{huddart_dataset_2024}. Coughs were collected using Android smartphones, which differed across countries~\cite{huddart_dataset_2024}. The TBscreen dataset was collected in Nairobi, Kenya in a controlled recording setting~\cite{sharma_tbscreen_2024}. Both passive coughs and forced coughs were obtained using three devices simultaneously. The Zambia CIDRZ dataset was originally used to establish the HeAR benchmark~\cite{baur_hear_2024}. It was collected from three clinical sites, Chawama, Chainda-South, and Kanyama. Participants were asked to produce four cough events, including three single coughs and one episode consisting of consecutive coughs, and were recorded using one professional audio recorder and three smartphone models simultaneously.

In this study, the released datasets were not used in their entirety. 
For CODA, we restricted analysis to recordings from Vietnam, Madagascar, and Tanzania, as these were the countries in which a single device was used consistently, enabling a cleaner assessment of device-related effects. Additionally, we used only the solicited coughs.
For TBscreen, passive and forced cough recordings were treated as separate resources. In their original study~\cite{sharma_tbscreen_2024}, passive and forced coughs were found to exhibit distinct characteristics, and models trained on passive coughs did not generalize well to forced coughs. We therefore kept the two cough types separate throughout the analysis. As passive cough represents the much larger portion of the dataset, passive cough recordings from TBscreen were used for training, whereas forced cough recordings were used only for validation. Additionally, for TBscreen passive, the number of coughs per subject is restricted to maximum 100 to limit overfitting to particular subjects and speed up the training process. This threshold was chosen based on the distribution of cough counts across subjects, where most subjects contributed between 100 and 200 coughs, while a small number of outliers exceeded 1000 coughs per subject. Capping at 100 therefore retains a representative sample for the majority of subjects while preventing a small number of high-count subjects from disproportionately influencing the training process.
For the Zambia dataset, the Chainda-South site data was removed from the analysis as this site had only one TB-positive subject. In addition, the subset recorded with the audio recorder represented only a small portion of the full dataset and was therefore used only for validation, and not for model training.

In addition to these dataset-level restrictions, we applied some audio preparation steps to obtain individual cough events in a comparable format across datasets. In CODA, the cough clips released had already been preprocessed into 0.5~s segments. However, the cough event was not consistently aligned within the segment, appearing near the beginning, center, or end, depending on the recording. Therefore, we applied an additional energy-based thresholding step to isolate the active portion of the cough. As a result, the final CODA cough segments used in this study were of variable duration rather than fixed 0.5~s length. A similar procedure was applied to TBscreen, whose original cough clips had been preprocessed into 1~s segments. The Zambia dataset required a different preparation procedure as the released audio files corresponded to full recording sessions rather than pre-segmented individual cough events. We therefore developed a custom cough extraction algorithm to first identify a rough interval of individual coughs, then the same energy-based thresholding method was applied to extract the coughs. As the same experimental session was often recorded simultaneously using multiple devices, detections from different devices could be cross-referenced to improve the final segmentation decision. During pre-processing of the Zambia data, we also observed that a small number of recordings appeared to be incorrectly labeled with respect to the subjects. For example, a recording was labeled as subject “03xxx" when, in fact, it was subject "01xxx”. This issue is particularly important because the nested CV procedure requires subject-level data splitting to avoid information leakage. As a validation step, we screened recordings from the same subject but from different devices by computing cross-correlations between the corresponding audio signals. Recordings with inconsistent cross-correlation patterns were flagged as potentially mismatched, manually checked and re-labeled.

Extended Data Table~\ref{tab:study_specific_subsets} summarizes the study-specific subsets used in the present analysis after dataset filtering, cough extraction, and validation control. We report, for each analyzed subset, the number of subjects, the number of cough events, and the TB+ rate at the subject-level based on publicly available data actually accessible for this study. These numbers do not always match those reported in the original dataset publications, because some datasets were only partially released. For example, in CODA, approximately half of the full dataset was made publicly available, while the remaining portion was withheld for continued benchmarking.

\subsection*{ML pipelines}
We included both classical ML and DL approaches. The aim is to assess whether poor cross-dataset generalization was primarily a limitation of the modeling framework or a consequence of the datasets themselves. The classical pipeline served as a transparent baseline: although simpler than DL, it can perform well when data are limited and offers greater interpretability in terms of feature selection and model behavior. The DL pipeline was included because it has become a standard approach for more complex audio classification tasks and may capture disease-related acoustic patterns that are not well represented by handcrafted features. In both pipelines, model development began with a search over data pre-processing choices, followed by model and representation selection, and concluded with external cross-dataset evaluation and post-analysis. Table~\ref{tab:pipeline_configurations} summarizes the pre-processing, augmentation, feature extraction, and model configurations explored for each pipeline.

\begin{table*}[h]
\centering
\caption{\textbf{Summary of explored configurations for the classical ML and deep learning pipelines.}
Hyperparameters of each augmentation method were determined by grid search.}
\label{tab:pipeline_configurations}
\renewcommand{\arraystretch}{1.35}
\setlength{\tabcolsep}{8pt}
\begin{tabular}{p{2.8cm} p{5.6cm} p{5.6cm}}
\toprule
& \textbf{Classical machine learning} & \textbf{Deep learning} \\
\midrule

\textbf{Filtering}
&
\multicolumn{2}{p{11.2cm}}{
  Bandpass: lower cut-off 10--500\,Hz, upper cut-off 4{,}000--8{,}000\,Hz.
} \\
\vspace{0.1cm}
\textbf{Augmentation}
&
\multicolumn{2}{p{11.2cm}}{
\vspace{0.1cm}
  Crop-and-pad, additive Gaussian noise, Brownian noise, tape-speed perturbation,
  amplitude scaling, pitch shifting, time stretching, circular time shifting~\cite{blankemeier_optimizing_2023}.
} \\

& \multicolumn{1}{l}{} &
\vspace{-0.3cm}
\begin{itemize}[noitemsep, topsep=2pt, leftmargin=*]
  \item Additionally: SpecAugment (frequency and time masking)
\end{itemize}
\\

\vspace{0.1cm}
\textbf{Features}
&
\vspace{0.1cm}
Feature selection via recursive backward elimination.
\begin{itemize}[noitemsep, topsep=2pt, leftmargin=*]
  \item CLAP embeddings
  \item Time-frequency domain features (Supplementary~F)
\end{itemize}
&
\begin{itemize}[noitemsep, topsep=2pt, leftmargin=*]
  \item Spectrogram
  \item Mel-spectrogram
  \item Scalogram (continuous wavelet transform)
\end{itemize}
\\
\vspace{0.1cm}
\textbf{Models}
&
\begin{itemize}[noitemsep, topsep=2pt, leftmargin=*]
  \item Random forest
  \item Logistic regression
  \item Gradient boosting
\end{itemize}
&
\begin{itemize}[noitemsep, topsep=2pt, leftmargin=*]
  \item ResNet-18, ResNet-34
  \item VGGish
  \item OPERA
  \item CLAP
\end{itemize}
\\

\bottomrule
\end{tabular}
\end{table*}

\subsubsection*{\textit{Classical ML pipeline}}
For the classical ML pipeline, model development began with the evaluation of candidate pre-processing strategies on the source dataset. The goal was to determine whether specific pre-processing and/or augmentation choices improved internal validation performance. The tested pre-processing methods included band-pass filtering, with lower cut-off frequencies ranging from 10 to 500~Hz and upper cut-off frequencies ranging from 4 to 8~kHz, covering the range commonly used in prior cough classification studies~\cite{sharma_tbscreen_2024, botha_detection_2018, pahar_automatic_2021}. In addition, several audio augmentation techniques were considered. Augmentations were applied at the level of individual audio samples after oversampling.

We then compared two feature families. The first consisted of conventional handcrafted acoustic descriptors extracted from the time and frequency domains. The full list of these features is provided in Supplementary~F. Specifically, frame-level features were computed from short-time analysis windows and then summarized across frames using their mean and standard deviation, resulting in a fixed-length representation for each cough recording. The second feature family consisted of embeddings extracted from a pretrained model, CLAP. CLAP (Contrastive Language--Audio Pretraining) is a large-scale representation-learning model trained to map audio signals and natural-language descriptions into a shared embedding space~\cite{elizalde_clap_2023}. In this study, we used only the pretrained acoustic encoder and treated its output as a general-purpose audio representation, without any additional pretrained or joint use of text information. The CLAP acoustic embedding is of dimension 1024.

For each feature representation, we compared a set of standard downstream classifiers on the source dataset. The models evaluated included logistic regression, random forests, and gradient boosting. These models were selected because they cover a range of linear and non-linear decision functions while remaining interpretable and well suited to tabular feature representations.

For both handcrafted time-frequency features and CLAP-derived features, model development on the source dataset followed a nested CV scheme. The outer loop split the data into five folds, and within the training portion of each outer fold, an inner four-fold CV was used to tune the classifier hyperparameters and perform feature selection with RFE. 
Additionally, feature scaling was fitted only on the training data of each fold and then applied unchanged to the corresponding test folds and to the external datasets.
Performance on the held-out outer folds, summarized as the ROC-AUC of subject-level predictions, served as the internal validation estimate and was used to select the final combination of preprocessing, feature set, and classifier for each source dataset. In addition to predictive performance, we evaluated the stability of feature selection by measuring the Jaccard similarity coefficient across CV folds and across datasets (see Supplementary~A).

Finally, we conducted a post-hoc analysis of the resulting feature spaces, both CLAP audio embedding and time-frequency domain features, to better understand the sources of poor generalization. In particular, we examined whether learned representations were organized more strongly by dataset, recording device, or subpopulation than by TB status. To quantify differences between groups, we computed pairwise distances using MMD and visualized the resulting distance structure in two dimensions using MDS. We also projected individual samples into t-SNE in order to inspect clustering patterns in the learned embeddings. Unlike the model-training experiments, all available cough samples were included in this analysis, including all CODA countries. The samples were used in their original form, without augmentation or resampling, so that the observed structure reflected the original feature distributions. The subgroups were defined according to location, recording device, or TB label. In particular, in CODA, the device and location were not independent, because each country analyzed was associated with a specific recording device or devices.

\subsubsection*{\textit{DL pipeline}}

For the DL pipeline, each cough recording was converted into a spectrogram-based image representation for input to neural network models. As in the classical pipeline, pre-processing and augmentation choices were first explored on the source dataset in order to identify configurations that improved internal validation performance. In addition to waveform-level augmentation, we also evaluated augmentation applied directly to the spectrogram representation using SpecAugment, in which time regions and frequency bands are randomly masked~\cite{blankemeier_optimizing_2023}. We further searched the spectrogram frequency range and the input representation itself, comparing standard spectrograms and mel-spectrograms. As the cough audios were segmented into arbitrary lengths, we padded the audios by repeating to the desired audio length depending on the model used.

As the available datasets remain limited relative to the data requirements of deep neural networks, we used pretrained audio models as feature-extraction backbones. A task-specific classification head was attached to each backbone to predict TB status from the learned representation. The evaluated backbones included ResNet-18, ResNet-34, VGGish, OPERA, and CLAP~\cite{he_deep_2015, hershey_cnn_2017, zhang_towards_2024, elizalde_clap_2023}. Brief descriptions of these models are provided in Supplementary~G.


Nested CV was used to control for overfitting during model training and evaluate model performance for model selection. Candidate configurations were defined through a grid search over pre-processing choices, spectrogram representations, backbone architectures, and model hyperparameters. For each candidate configuration, training was performed within the inner CV loop: the model was fitted on the inner training split, while the corresponding inner validation split was used to monitor validation performance and determine early stopping. The performance of each candidate configuration was then summarized by averaging the results across the outer folds.
We report the configuration with the highest average outer-fold results within the source dataset. The models were subsequently evaluated under external validation on the remaining datasets. Performance was summarized using ROC-AUC.

\subsection*{Analysis of location-related or device-related bias in CODA}

We further investigated how models could exploit non-disease structure using the CODA dataset.
This analysis was motivated by the observation that CODA combines recordings from multiple countries, each with its own TB+ rate and recording setup. Under such conditions, a model trained for TB classification may partially learn country-, site-, or device-related information if these factors are correlated with disease status.

We compared two modeling settings. The first setting was designed to closely reproduce the best-performing CODA TB DREAM Challenge pipeline~\cite{jaganath_accelerating_2025}. This pipeline used spectrogram representations computed with a Hann window of length 1024 samples and a hop size of 64 samples, restricted to the frequency range between 50 and 15000~Hz. These spectrograms were used as input to a ResNet-34 classifier. The original pipeline used both solicited cough recordings and longitudinal cough recordings, and did not apply additional audio pre-processing steps such as filtering or audio augmentation. In the present study, for simplicity, we only used the solicited cough recordings to replicate the pipeline. Resampling was performed to balance only the TB+ and TB- classes.

The second setting was designed as a more conservative pipeline to reduce the influence of acquisition-related confounding. Similarly, only solicited coughs were used. In this setting, we used a VGGish-based model with spectrogram inputs restricted to frequencies below 8,000~Hz. This upper frequency limit was chosen because most cough-relevant acoustic information is expected to be contained within this range~\cite{korpas_analysis_1996, sharma_tbscreen_2024, botha_detection_2018, pahar_automatic_2021}, while higher-frequency components may contain device- or environment-specific artifacts. Additionally, cough segmentation was applied before feature extraction so that the model focused on the active cough event rather than surrounding silence or background sound. We also used a stricter resampling strategy that balanced the training data jointly by TB label and country source. This was intended to deny model the disease rate prior information specific to each country.

For both settings, we evaluated whether the resulting predicted TB probabilities contained information about the country source and the prevalence of the disease at the country-level. We compared the distribution of predicted TB probabilities across groups defined by country and TB label. 
To quantify the association between model output and country-level prevalence, we computed the linear regression slope and $R^2$ between the mean predicted TB probability per country and the country-level disease rate, calculated separately for TB- and TB+ cases. The regression slope captures the magnitude of the association, indicating how much the mean predicted probability changes per unit increase in disease rate, while $R^2$ describes the proportion of variance in the mean model output that is explained by country-level prevalence alone. 
The observed high $R^2$ combined with a steep slope in the challenge-replication approach indicated that the model output was strongly driven by the prevalence structure associated with the country, while the lower $R^2$ and flat slope in the conservative approach suggested that the predictions were less influenced by the country of origin.

This analysis was not intended to evaluate a deployable TB classifier. Instead, it was used as a diagnostic experiment to assess whether resampling, modeling and, pre-processing choices could reduce the influence of acquisition-related or prevalence-related shortcuts in a heterogeneous multi-country cough dataset.

\subsection*{Analysis of device-related bias in the Zambia dataset}

The location-related bias analysis in CODA provides evidence that cough-based classifiers exploit non-disease structure when TB status is correlated with acquisition-related factors. However, in CODA, several sources of heterogeneity are entangled, including country, recording protocol, clinical setting, disease prevalence, and recording device. Therefore, although the analysis demonstrates the presence of acquisition-related bias, it cannot isolate the contribution of the recording device alone.

The difference in the recording device is important in acoustic ML studies, as it can have a direct effect on the measured signal. Microphones and mobile devices differ in frequency response, gain control, noise suppression, compression, and other signal-processing characteristics, all of which can alter the spectral and temporal properties of recorded cough sounds. Previous work in cough detection and broader audio classification has shown that device mismatch can reduce model robustness and external generalization, particularly when models are evaluated on devices not represented during training~\cite{barata_towards_2019, yang_device-invariant_2026}. Similar device-domain effects are also well recognized in acoustic scene classification and other machine-hearing tasks, where recording-device differences are a major source of domain shift~\cite{masoudian_domain_2023}. Thus, while population and site effects may also contribute to poor generalization, device variation is a plausible and important source of instability in cough-based models.

To examine the effect of recording device more directly, we performed an additional set of experiments using the Zambia dataset. This dataset is particularly suitable for this purpose because cough recordings were collected from the same clinical sites using multiple recording devices simultaneously. This structure allowed us to evaluate device-related generalization while reducing, although not completely eliminating, confounding by population and recruitment setting.
We considered all four recording devices: an audio recorder, Phone A (Pixel 3a, lower-tier), Phone B (Galaxy A12, mid-tier), and Phone C (Galaxy A22, higher-tier). The analysis was restricted to the Chawama and Kanyama sites, as the Chainda-South site had only one TB+ subject. Unlike the main modeling pipeline, the audio-recorder subset was included here as both a possible training and testing device. All experiments in this analysis adopted the DL pipeline.

We considered two experimental settings, where within-device evaluation is done by nested CV. In particular, for each outer fold, the training subjects were first identified and then external-device evaluation was conducted only on recordings of subjects that were not included in the corresponding training split. In the first setting, a DL model was trained using recordings from a single device and evaluated both on held-out recordings from the same device and on recordings from the remaining devices. This experiment tested whether a model trained on one device retained performance when applied to audio collected with a different device. In the second setting, the model was trained using recordings from three devices and evaluated on the fourth device, which was not used during training. This experiment tested whether exposure to multiple devices during training improved generalization to an unseen recording device.

Together, these experiments addressed two related questions. First, we asked whether recording-device mismatch leads to a measurable decrease in TB classification performance compared with within-device evaluation. Second, we asked whether training on a more device-diverse dataset improves external-device performance. By comparing results within and across-devices in both settings, this analysis provided a more targeted assessment of the extent to which device variation contributes to the generalization failures observed in cough-based TB classification.

\subsection*{Clinical-variable baseline}

In addition to cough-based models, we evaluated whether routinely collected clinical variables could support TB classification across datasets. This analysis served two purposes. First, it provided a non-acoustic baseline for comparison with the cough-based models. Second, it allowed us to examine whether the generalization problem observed for cough acoustics was specific to audio recordings, or whether similar limitations also appeared when using structured clinical information. If clinical models generalize better than acoustic models, this would suggest that part of the difficulty lies in acquisition-related variability in audio data. Conversely, if clinical models also show poor external performance, this would indicate that differences in study population, recruitment criteria, symptom distributions, and disease prevalence also contribute substantially to the generalization problem. This baseline should not be interpreted as a replacement for microbiological diagnosis, but rather as a means to contextualize the acoustic results.

We restricted this analysis to variables that were available across the datasets in a sufficiently comparable form. The shared clinical variables included age, sex, smoking status, HIV status, previous TB history, cough duration, fever, weight loss, and night sweats. These variables capture known clinical and epidemiological risk factors for TB and are commonly available in screening contexts. Other variables were available only in individual datasets and may have been predictive within those datasets. For example, our preliminary analyses suggested that heart rate carried useful information in some settings. However, these dataset-specific variables were not included in the cross-dataset baseline because they were not consistently available across all datasets.

Samples with missing clinical-variable values or missing TB labels were removed from this analysis. After this filtering step, the clinical-variable dataset contained 1,823 participants from Zambia, including 205 TB+ and 1,618 TB- participants; 1,037 participants from CODA, including 282 TB+ and 755 TB- participants; and 125 participants from TBscreen, including 89 TB+ and 36 TB- participants. Because this analysis used participant-level clinical variables rather than cough recordings, Zambia was not split by recording device, TBscreen was not split into passive and forced cough subsets, and CODA was not split by countries.

Continuous variables, including age and cough duration, were standardized before model fitting. Standardization parameters were computed on the training data only and then applied to the test data. Categorical or binary variables were encoded in a consistent format across datasets where possible. We used logistic regression as a simple and interpretable classifier, as the purpose of this analysis was to determine whether basic clinical information can generalize.

Clinical-variable models were trained and evaluated using the same dataset-wise validation principle as the acoustic models. In each experiment, one dataset was used as the source dataset for model development, and the remaining datasets were used for external validation. Within the source dataset, model development was performed using nested CV: the outer loop was used to estimate internal validation performance, while the inner loop was used for logistic-regression hyperparameter tuning. The final selected model was then evaluated on the external datasets that were not used during training or tuning. This design allowed us to compare the acoustic and clinical models under the same external-validation framework.

We also assessed the importance of the clinical variables in two complementary ways. First, we used the fitted model coefficients as a direct measure of feature association. Positive coefficients indicate that larger feature values were associated with a higher predicted probability of TB, whereas negative coefficients indicate an association with a lower predicted probability of TB. Second, we computed permutation importance using the trained model and the corresponding testing fold. For each feature, its values were randomly permuted and the decrease in ROC-AUC after permutation was used as an empirical estimate of importance. A larger decrease indicates that the model relied more strongly on that feature, whereas values close to zero, or negative values, indicate little contribution to predictive performance. Because the clinical-variable pipeline was evaluated using five outer CV folds, feature-importance estimates were computed separately for each fold and then summarized across folds for each training dataset.

\bmhead{Data availability}
All data sets analyzed in this study are publicly available. No new data were collected for this work.

The CODA data set is available through the Synapse platform under project
ID \texttt{syn31472953} (\url{https://www.synapse.org/Synapse:syn31472953}),
released in the context of the CODA TB DREAM Challenge~\cite{huddart_dataset_2024}. 

The TBscreen data set is available at
\url{https://zenodo.org/records/10431329}, as described in
the original publication~\cite{sharma_tbscreen_2024}.

The Zambia data set corresponds to the CIDRZ TB cough cohort collected by the
Centre for Infectious Disease Research in Zambia and released in connection
with the HeAR benchmark~\cite{baur_hear_2024}. It is available at
\url{https://www.kaggle.com/datasets/googlehealthai/google-health-ai}.

\bmhead{Code availability}

 All source code used in the study is publicly available at \url{https://github.com/esl-epfl/tb-cough-generalization}.

\clearpage
\bibliography{sn-bibliography}
\clearpage

\bmhead{Author contributions}
W.Z. implemented the experiments, performed the analyses, and wrote the initial manuscript. All authors contributed to the conceptualization of the study, designed the experiments, discussed the results, and reviewed the manuscript. All authors read and approved the final manuscript.

\bmhead{Funding}

This research was supported in part by the Wyss Center for Bio and Neuro Engineering: Lighthouse Noninvasive Neuromodulation of Subcortical Structures, in part by the Swiss State Secretariat for Education, Research, and Innovation (SERI) through the SwissChips research project, and in part by the ETH Domain Open Research Data (ORD) Program under the 8th `Contribute' Call. This work has additionally been partially supported by the grants RYC2021-032853-I and PID2024-156173OA-I00, funded by MCIN/AEI/10.13039/501100011033 and by the European Union.

\bmhead{Competing interests}
The authors declare no competing interests.

\clearpage
\setcounter{figure}{0}
\setcounter{table}{0}
\renewcommand{\thefigure}{\arabic{figure}}
\renewcommand{\thetable}{\arabic{table}}
\renewcommand{\figurename}{Extended Data Fig.}
\renewcommand{\tablename}{Extended Data Table}

\section*{Extended Data Figure 1 Cross-dataset performance with 95\% CI}

\begin{figure}[H]
    \centering
    \begin{subfigure}[t]{0.49\textwidth}
        \centering
        \includegraphics[width=\linewidth]{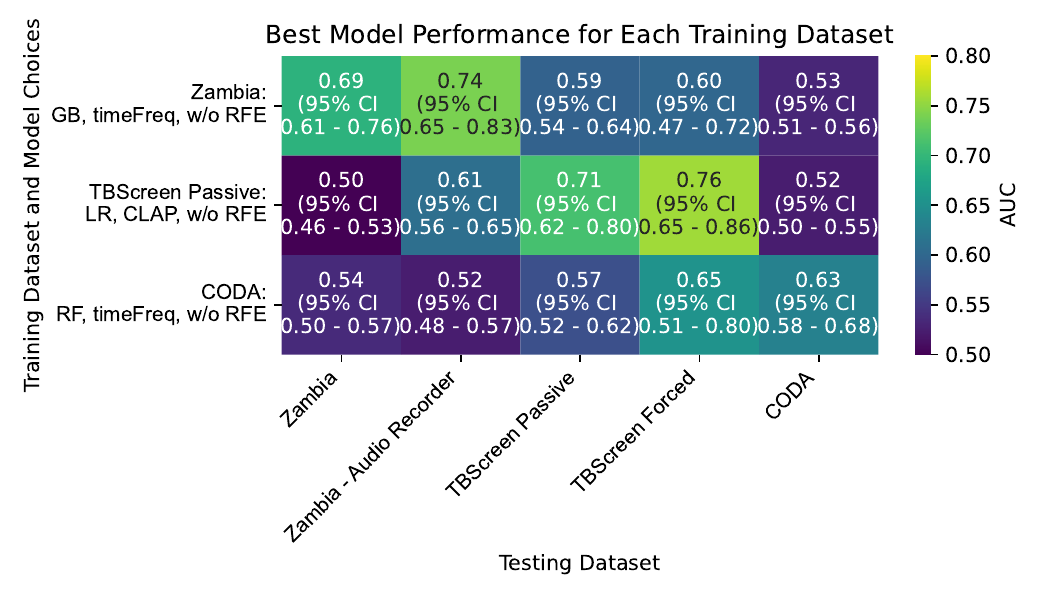}
        \caption{Classical ML pipeline}
    \end{subfigure}
    \hfill
    \begin{subfigure}[t]{0.49\textwidth}
        \centering
        \includegraphics[width=\linewidth]{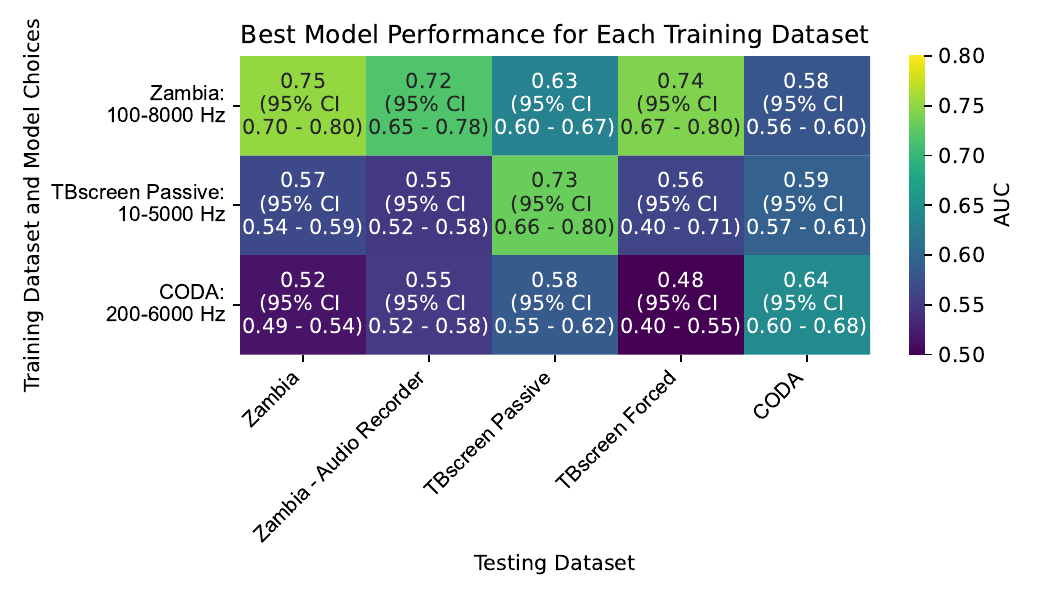}
        \caption{DL pipeline}
    \end{subfigure}
    \vspace{0.8em}
    \begin{subfigure}[t]{0.5\textwidth}
        \centering
        \includegraphics[width=0.80\linewidth]{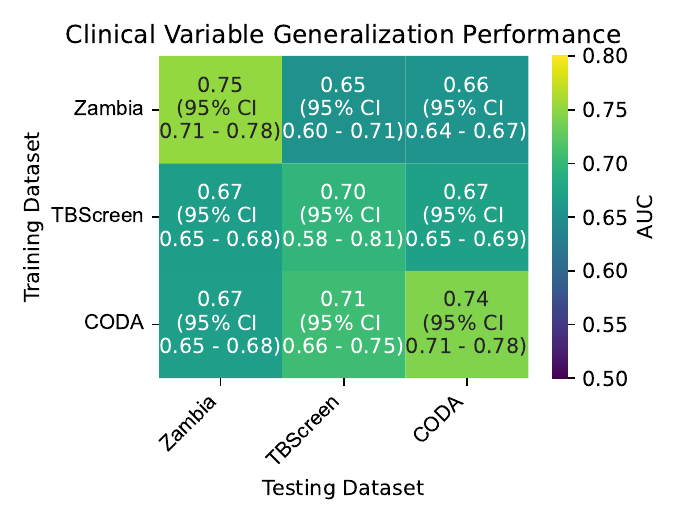}
        \caption{Clinical-variable baseline}
    \end{subfigure}
    \caption{
    \textbf{Cross-dataset performance with 95\% confidence intervals.}
    Same models and evaluation as in Fig.~\ref{fig:best_models}. For each source dataset, the model configuration was selected using within-dataset validation and then evaluated on all target datasets. Values show subject-level ROC-AUC with 95\% CI estimated by participant-level bootstrap. The confidence intervals confirm the pattern in the main analysis: acoustic models generalize inconsistently across datasets, whereas the clinical-variable baseline transfers more stably.
    }
    \label{fig:best_models_ci}
\end{figure}

Extended Data Fig.~\ref{fig:best_models_ci} reproduces the cross-dataset results of Fig.~\ref{fig:best_models}, reporting 95\% CI instead of the variability across outer folds, to show the uncertainty of each internal and external estimate. Confidence intervals were estimated by participant-level bootstrap, as described in Methods.

\section*{Extended Data Figure 2 Classical ML performance across configurations}

\begin{figure}[H]
    \centering
    \includegraphics[width=\linewidth]{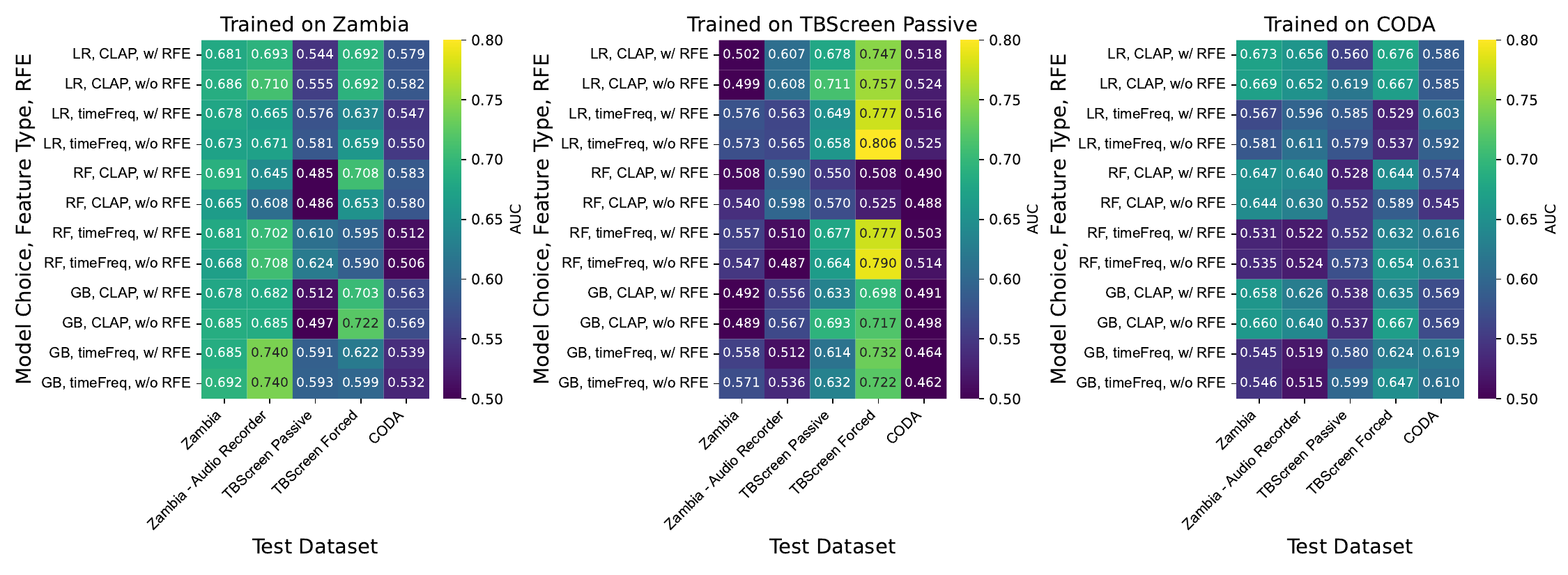}
    \caption{
    \textbf{Classical ML performance across training datasets, testing datasets, classifiers, feature representations, and feature-selection settings.}
    Each heatmap corresponds to one source training dataset. Rows show combinations of classifier, feature type, and RFE; columns show testing datasets. Performance is reported as subject-level ROC-AUC. Overall, performance varied more strongly by training and testing dataset than by classifier choice, feature representation, or use of feature selection.
    }
    \label{fig:classical_ml_grand_performance}
\end{figure}

\begin{figure}[H]
    \begin{subfigure}[t]{0.98\textwidth}
        \centering
        \includegraphics[width=\linewidth]{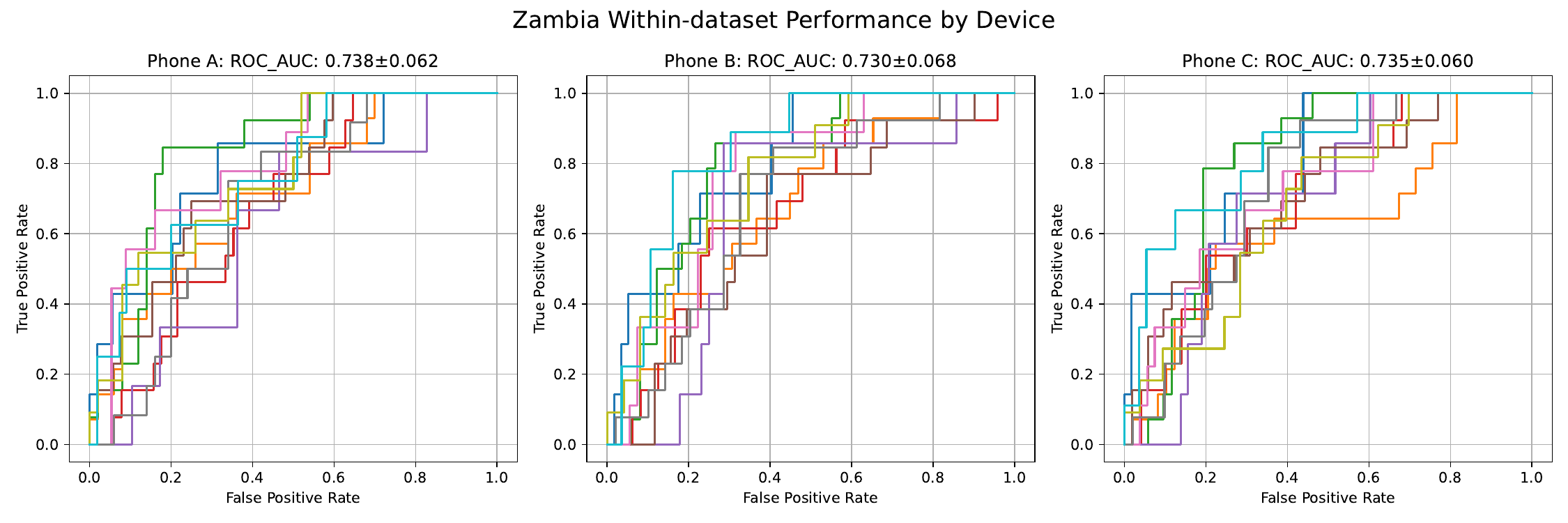}
        \caption{By recording device.}
    \end{subfigure}
    \vspace{0.8em}

    \begin{subfigure}[t]{0.98\textwidth}
        \centering
        \includegraphics[width=0.66\linewidth]{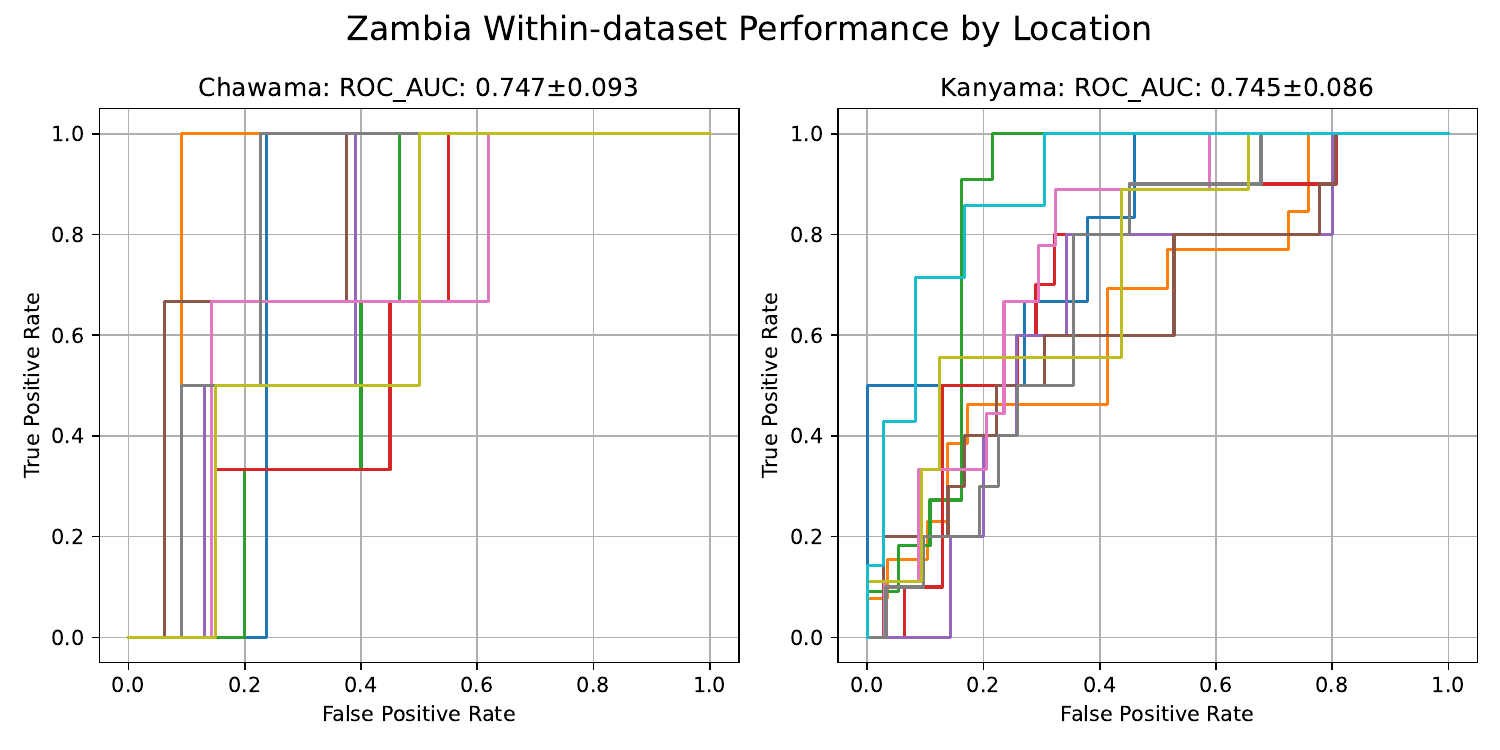}
        \caption{By location.}
    \end{subfigure}
    \caption{\textbf{Within-dataset DL performance for the Zambia dataset, stratified by recording device and by location.}
    Subject-level ROC-AUC for the DL pipeline evaluated within Zambia, shown separately for each recording device and for each recruitment site.}
    \label{fig:Zambia_spec}
\end{figure}

\begin{figure}[H]
    \centering
    \includegraphics[width=\linewidth]{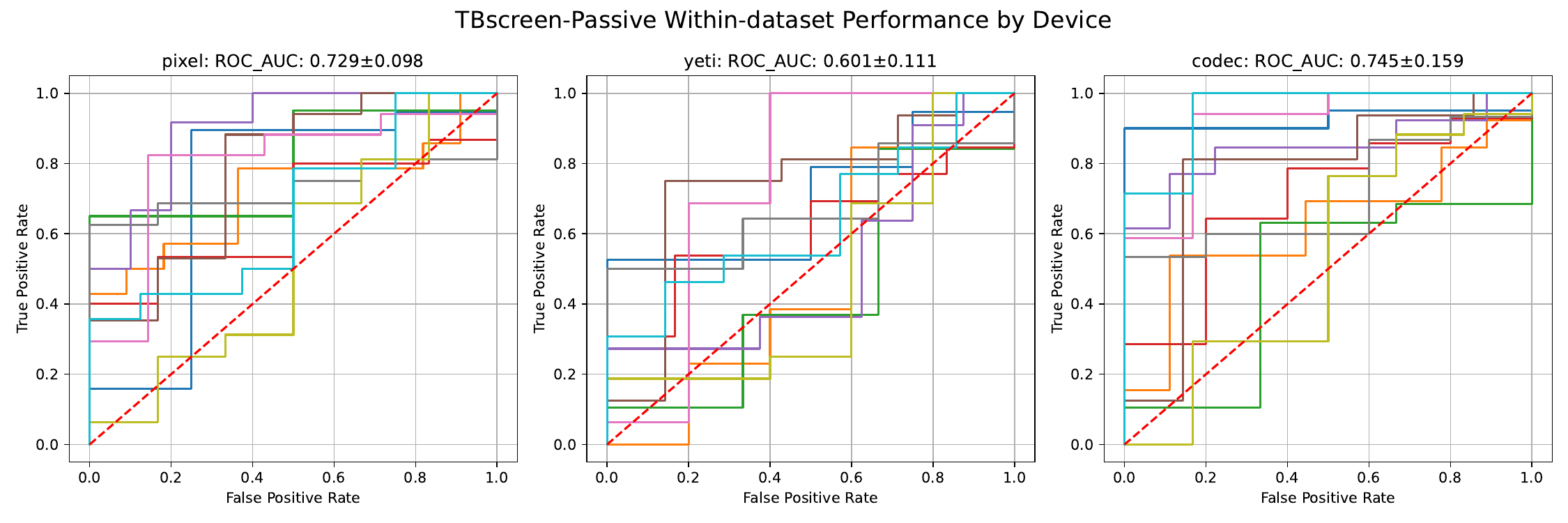}
    \caption{
    \textbf{Within-dataset performance analysis for TBscreen dataset stratified by device.}
    Subject-level ROC-AUC for the DL pipeline evaluated within TBscreen passive cough, shown separately for each recording device. Performance is similar between Pixel and codec, but is lower when using yeti.}
    \label{fig:tbscreen_spec_device}
\end{figure}

\begin{figure}[H]
    \centering
    \includegraphics[width=\linewidth]{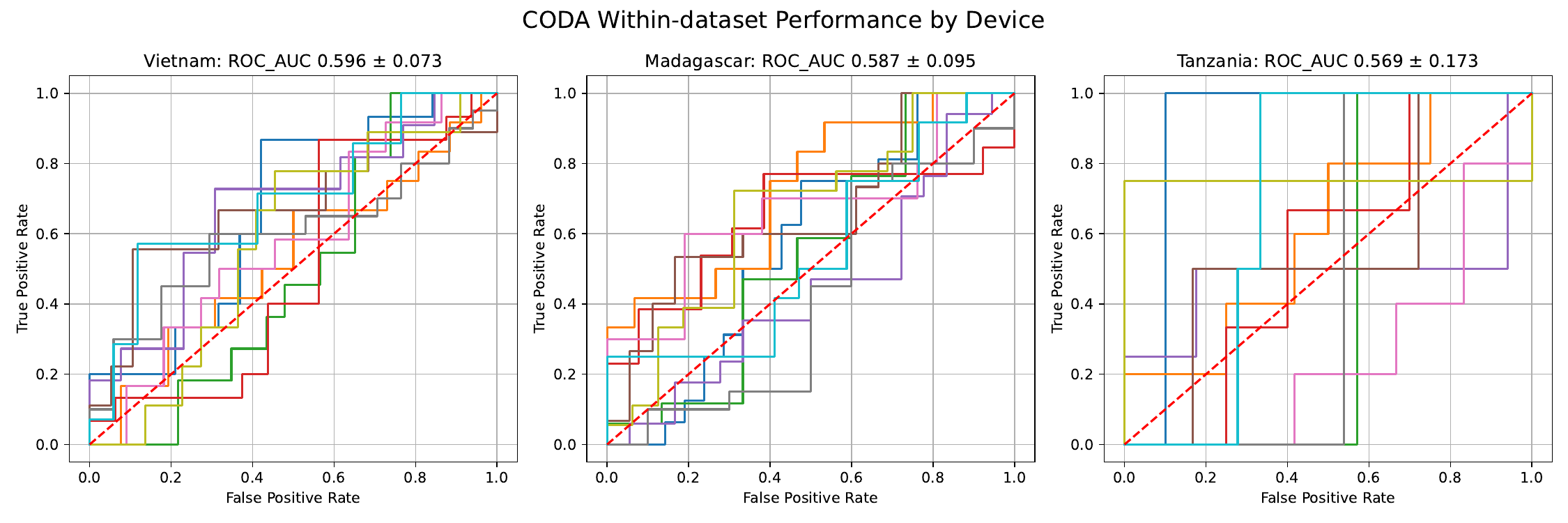}
    \caption{
    \textbf{Within-dataset performance analysis for CODA dataset stratified by device.}
    Subject-level ROC-AUC for the DL pipeline evaluated within CODA, shown separately for each recording device, each of which also corresponds to a different country. Performance variability is too large to draw any conclusion.}
    \label{fig:coda_spec_device}
\end{figure}

\begin{figure}[H]
    \centering
    \includegraphics[width=0.49\linewidth]{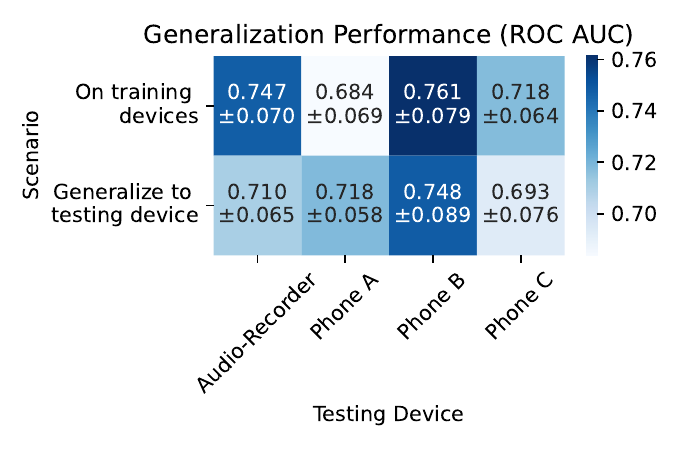}
    \hfill
    \includegraphics[width=0.49\linewidth]{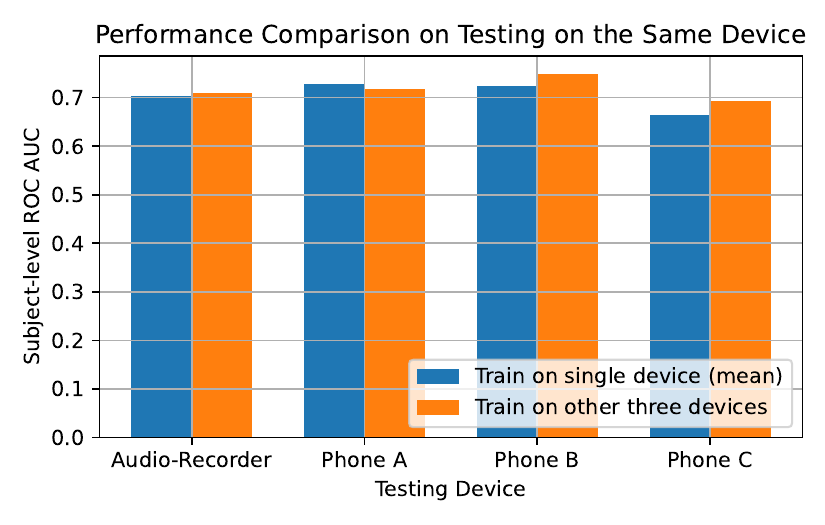}
    \caption{
    \textbf{Leave-one-device-out transfer with sample-count-matched training.}
    Values are subject-level ROC-AUC.
    The setup is the same as in Fig.~\ref{fig:zambia_device}, except that the training data from the three remaining devices were resampled to match the number of samples in the single-device setting. Multi-device training still improves transfer to the held-out device, indicating that the gain reflects device diversity rather than training set size.
    }
    \label{fig:device_multi_matched}
\end{figure}

Extended Data Fig.~\ref{fig:classical_ml_grand_performance} summarizes classical ML performance across all combinations of source dataset, external validation dataset, feature representation, classifier, and feature-selection setting. No single modeling choice consistently dominated across datasets. Differences between classifiers, feature families, and feature-selection settings were generally modest compared with the differences between training and testing datasets. This indicates that the main limitation of the classical pipeline was not the choice of downstream classifier, feature descriptor, or feature-selection strategy, but the mismatch between datasets, consistent with the best-model results reported in the main text.

\section*{Extended Data Figure 3 Stratified within-dataset performance analysis of DL approach of the Zambia dataset}

For the Zambia dataset, performance is even across both recording devices and locations, which indicates that the within-dataset Zambia result is not driven by a particular device or site.

\section*{Extended Data Figure 4 Stratified within-dataset performance analysis of DL approach of the TBscreen passive cough dataset}

For TBscreen passive cough, performance differs moderately across recording devices, pointing to some device sensitivity even within a single dataset. 

\section*{Extended Data Figure 5 Stratified within-dataset performance analysis of DL approach of the CODA dataset}

For CODA, the variability of the performance itself is too big to draw meaningful conclusions.

\section*{Extended Data Figure 6 Leave-one-device-out transfer with sample-count-matched training}

To test whether the benefit of multi-device training could be explained by the larger amount of training data, we repeated the leave-one-device-out analysis under a sample-count-matched condition. For each held-out test device, the training data from the three remaining devices were resampled so that the total number of training samples matched that of the corresponding single-device setting. All other settings, including the pipeline configuration and evaluation procedure, were kept identical to the main analysis.

\section*{Extended Data Figure 7 Clinical-variable feature importance}

\begin{figure}[H]
    \centering
    \includegraphics[width=\linewidth]{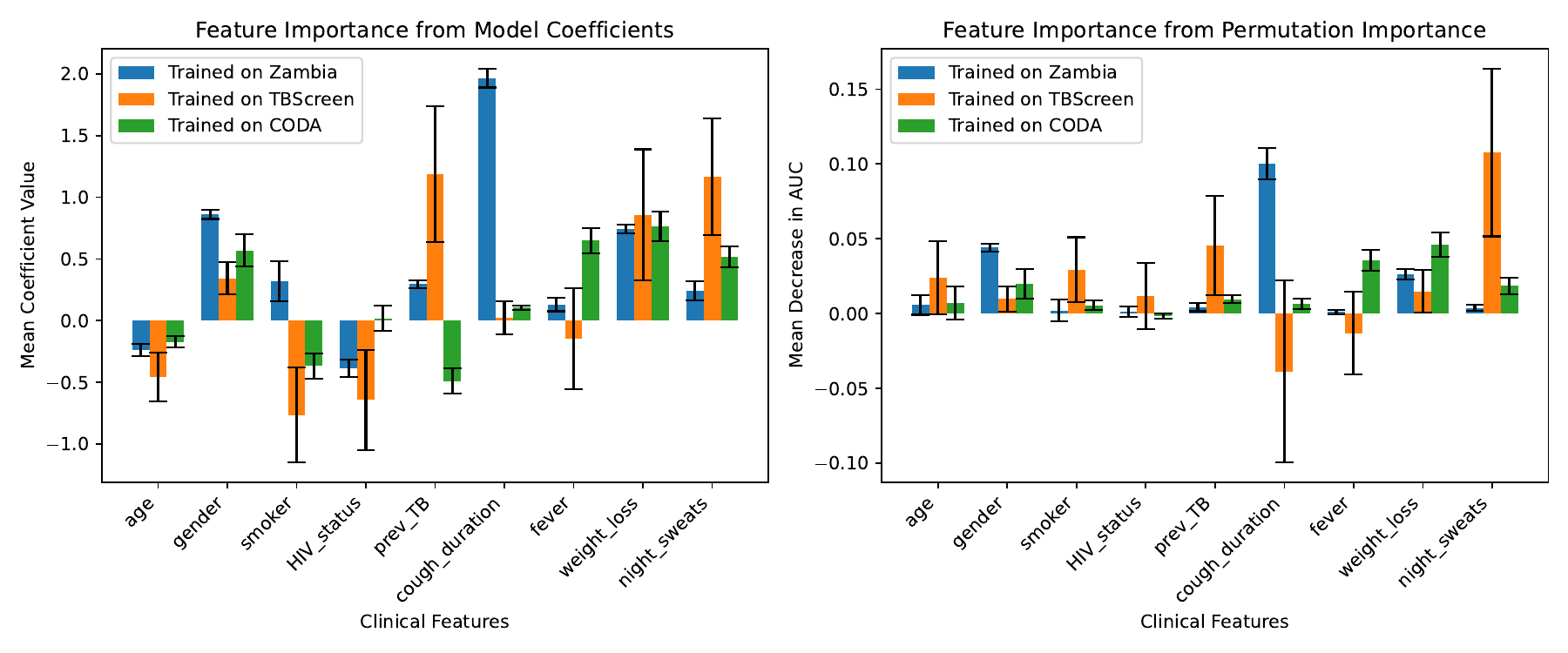}
    \caption{
    \textbf{Feature importance of the clinical-variable baseline.}
    The left panel shows mean logistic-regression coefficients across outer CV folds for each training dataset. Positive coefficients indicate an association with higher predicted TB probability, while negative coefficients indicate an association with lower predicted TB probability. The right panel shows permutation importance, measured as the mean decrease in ROC-AUC after randomly permuting each feature in the testing fold.
    }
    \label{fig:clinical_feature_importance}
\end{figure}

We examined which clinical variables contributed to the predictions of the clinical-variable baseline, using both model coefficients and permutation importance (Extended Data Fig.~\ref{fig:clinical_feature_importance}). The two analyses showed both consistent patterns and dataset-specific differences. Some associations were relatively stable across training datasets. Weight loss was positively associated with TB risk in all three datasets and contributed positively under permutation analysis, and night sweats showed positive coefficients across datasets, though its empirical importance varied. These patterns are consistent with the role of systemic symptoms in clinical TB screening. Other variables were more dataset-dependent. Smoking status was positively associated with TB prediction in the Zambia-trained model but negatively in the TBscreen- and CODA-trained models; cough duration was highly influential in the Zambia-trained model yet contributed little in CODA and was unstable in TBscreen; and previous TB history was positive in the Zambia- and TBscreen-trained models but negative in CODA. These differences indicate that even when clinical-variable models generalize better than acoustic models, the decision rules they learn are not identical across datasets. 

\section*{Extended Data Table 1 Study-specific dataset subsets}

The released datasets were not analyzed in their entirety. For each dataset, we selected the subsets best suited to the current study. Extended Data Table~\ref{tab:study_specific_subsets} summarizes the resulting subsets, reporting for each one the recording device, the number of subjects, the number of cough events, the subject-level TB+ rate, and whether the subset was used for training and validation or for validation only. The reported counts reflect the data publicly accessible and may differ from the figures in the original dataset publications.

\begin{table}[h]
\centering
\caption{\textbf{Summary of the study-specific dataset subsets analyzed in this work}. Disease rate is reported at the subject level.}
\label{tab:study_specific_subsets}
\renewcommand{\arraystretch}{1.2}
\setlength{\tabcolsep}{5pt}
\begin{tabular}{p{1.4cm} p{2.1cm} p{3.7cm} p{1.9cm} p{1.6cm} p{1.5cm} p{2.3cm}}
\hline
\textbf{Dataset} & \textbf{Subgroup} & \textbf{Device} & \textbf{Number of subjects} & \textbf{Number of cough events} & \textbf{TB+ rate (\%)} & \textbf{Usage} \\
\hline

CODA & Vietnam & OPPOA54 & 161 & 1791 & 40 & Train / Vali \\
CODA & Madagascar & Motorola G9 play & 157 & 1293 & 48 & Train / Vali \\
CODA & Tanzania & Nokia 3.4 Ta-1288
 & 87 & 1004 & 16 & Train / Vali \\

TBscreen & Passive cough & Pixel & 118 & 3762 & 70 & Train / Vali \\
TBscreen & Passive cough & Boundary microphone & 114 & 3370 & 72 & Train / Vali \\
TBscreen & Passive cough & Condenser microphone & 106 & 2606 & 74 & Train / Vali \\
TBscreen & Forced cough & Pixel & 38 & 410 & 87 & Vali only \\
TBscreen & Forced cough & Boundary microphone & 34 & 366 & 82 & Vali only \\
TBscreen & Forced cough & Condenser microphone & 29 & 234 & 83 & Vali only \\

Zambia & Chawama & Phone A & 111 & 687 & 9 & Train / Vali \\
Zambia & Chawama & Phone B & 102 & 595 & 10 & Train / Vali \\
Zambia & Chawama & Phone C & 109 & 635 & 9 & Train / Vali \\
Zambia & Chawama & Audio recorder & 90 & 520 & 10 & Vali only \\
Zambia & Kanyama & Phone A & 207 & 1158 & 22 & Train / Vali \\
Zambia & Kanyama & Phone B & 214 & 1229 & 22 & Train / Vali \\
Zambia & Kanyama & Phone C & 214 & 1205 & 22 & Train / Vali \\
Zambia & Kanyama & Audio recorder & 120 & 741 & 21 & Vali only \\

\hline
\end{tabular}
\end{table}

\end{document}


\begin{appendices}

\section{Recursive feature elimination for the classical ML pipeline}\label{app:classical_ml_grand_performance}

\begin{figure}[h]
    \centering
    \includegraphics[width=0.6\linewidth]{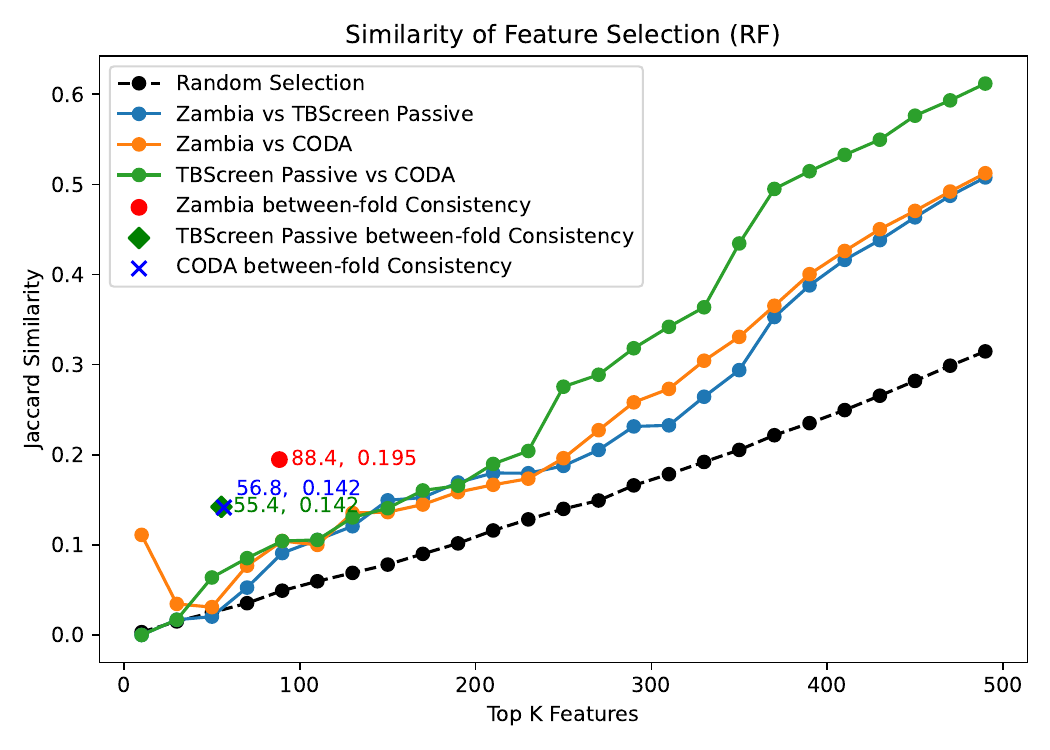}
    \caption{
    \textbf{Feature-selection consistency for CLAP embeddings with random forest classifiers.}
    Jaccard similarity was used to compare selected feature subsets across source datasets and across CV folds. The selected CLAP dimensions showed greater overlap than random selection, suggesting that RFE identified some stable embedding dimensions in this setting. For between-fold consistency, the mean number of selected features and the mean Jaccard similarity between folds were plotted as scatter points and marked on the plot.
    }
    \label{fig:clap_rf_feature_selection_similarity}
\end{figure}

We examined whether recursive feature elimination (RFE) improved performance or stability by comparing the mean and standard deviation of ROC-AUC with and without feature selection for each feature representation and classifier. In most settings, feature selection did not clearly improve average performance or stability, the only exception being CLAP embeddings combined with random forest classifiers. For this combination, we assessed selection consistency using the Jaccard similarity coefficient, both across source datasets and across folds within the nested CV pipeline. As shown in Fig.~\ref{fig:clap_rf_feature_selection_similarity}, the overlap between selected CLAP dimensions exceeded that of random feature selection, and between-fold consistency was likewise above the random baseline. Consistency was generally higher than random for CLAP embeddings with random forest or gradient boosting classifiers, but weaker for CLAP-based logistic regression. For the lower-dimensional handcrafted time-frequency features, RFE often stopped after removing few features and the selected subsets were not consistently more stable than random. The repeatedly selected CLAP dimensions did not admit a reliable clinical interpretation, since CLAP embeddings are learned representations that do not map straightforwardly onto individual acoustic properties; exploratory interpretation~\cite{zhang_transformation_2025} did not yield consistent patterns. Overall, feature selection can improve the compactness of some CLAP-based models, particularly with tree-based classifiers, but it does not resolve the broader problem of limited external generalization.

\section{Representation visualization of classical ML pipeline}\label{app:representation_visual_classical}
\begin{figure}[h]
    \centering
    \begin{subfigure}[t]{0.32\textwidth}
        \centering
        \includegraphics[width=\linewidth]{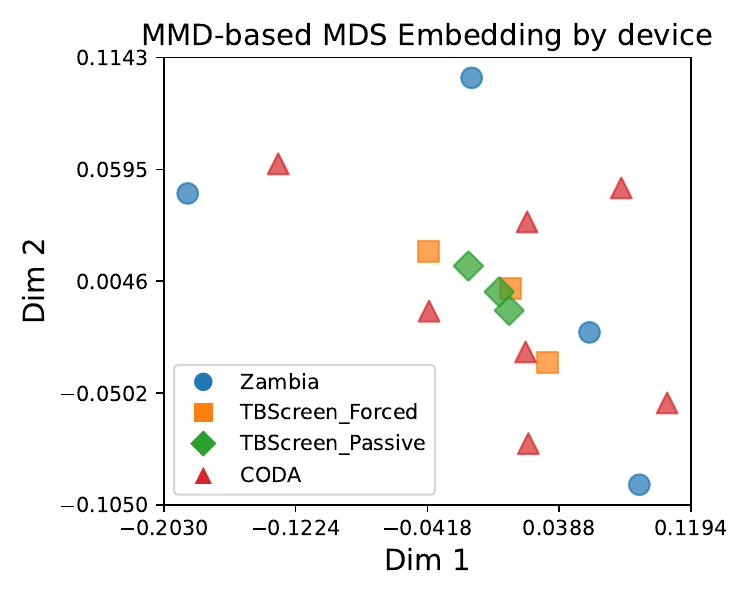}
        \caption{Device}
        \label{fig:mmd_timefreq_device}
    \end{subfigure}
    \hfill
    \begin{subfigure}[t]{0.32\textwidth}
        \centering
        \includegraphics[width=\linewidth]{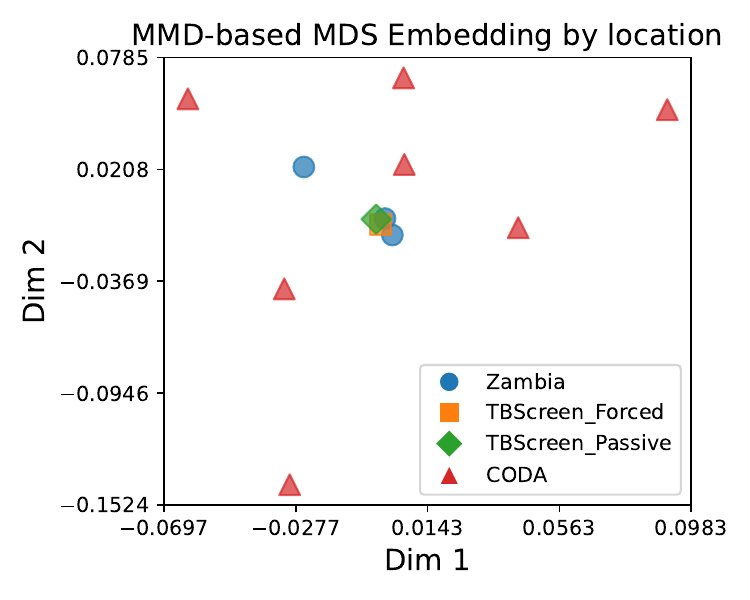}
        \caption{Location}
        \label{fig:mmd_timefreq_location}
    \end{subfigure}
    \hfill
    \begin{subfigure}[t]{0.32\textwidth}
        \centering
        \includegraphics[width=\linewidth]{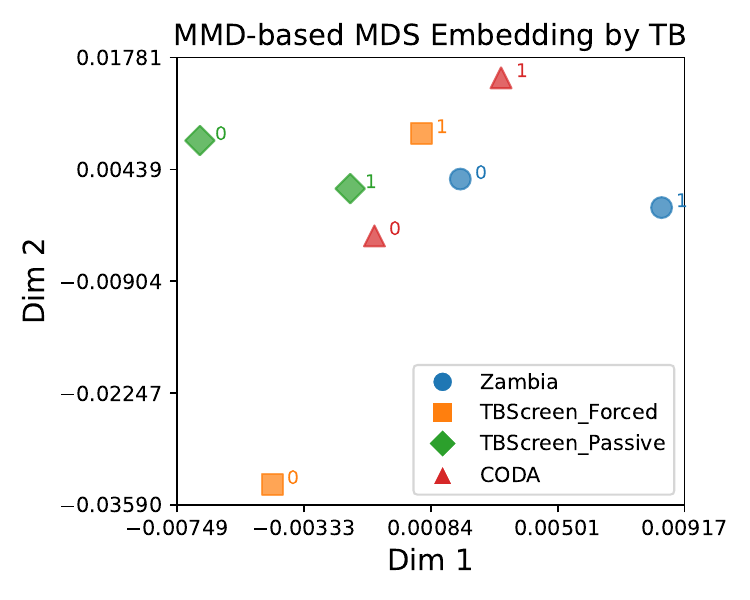}
        \caption{TB label}
        \label{fig:mmd_timefreq_tb}
    \end{subfigure}

    \caption{
    \textbf{MMD-based MDS visualization of subgroup structure in handcrafted time-frequency features.}
    As with CLAP embeddings, the time-frequency feature space showed stronger separation by recording device than by TB label. However, the separation by dataset domain is generally weaker than CLAP embeddings.
    }
    \label{fig:mmd_timefreq_subgroups}
\end{figure}

\begin{figure}[h]
    \centering
    \includegraphics[width=0.9\linewidth]{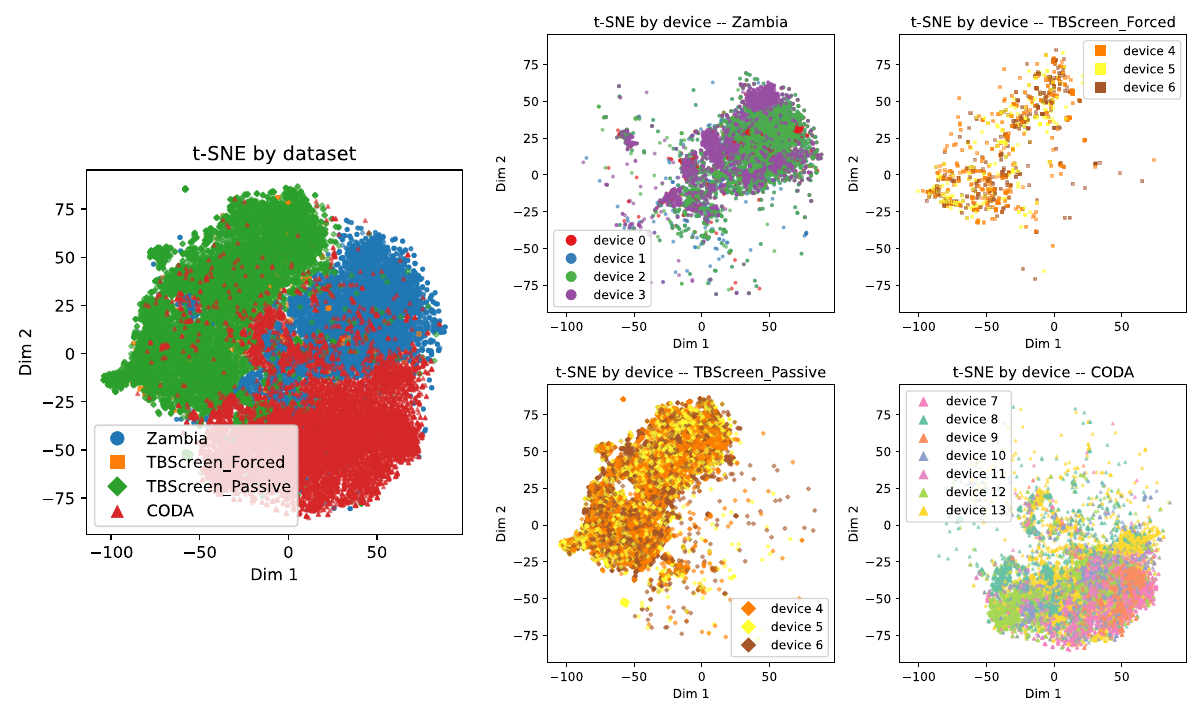}
    \caption{
    \textbf{t-SNE visualization of sample-level feature structure in CLAP embedding by dataset and recording device.}
    Individual cough samples were projected into two dimensions using t-SNE. Samples showed strong organization by dataset source and recording device. This acquisition-related clustering was more prominent than separation by TB label, supporting the conclusion that dataset and device structure dominate the feature spaces used by the classical ML models.
    }
    \label{fig:tsne_device_clap}
\end{figure}

\begin{figure}[h]
    \centering
    \includegraphics[width=0.9\linewidth]{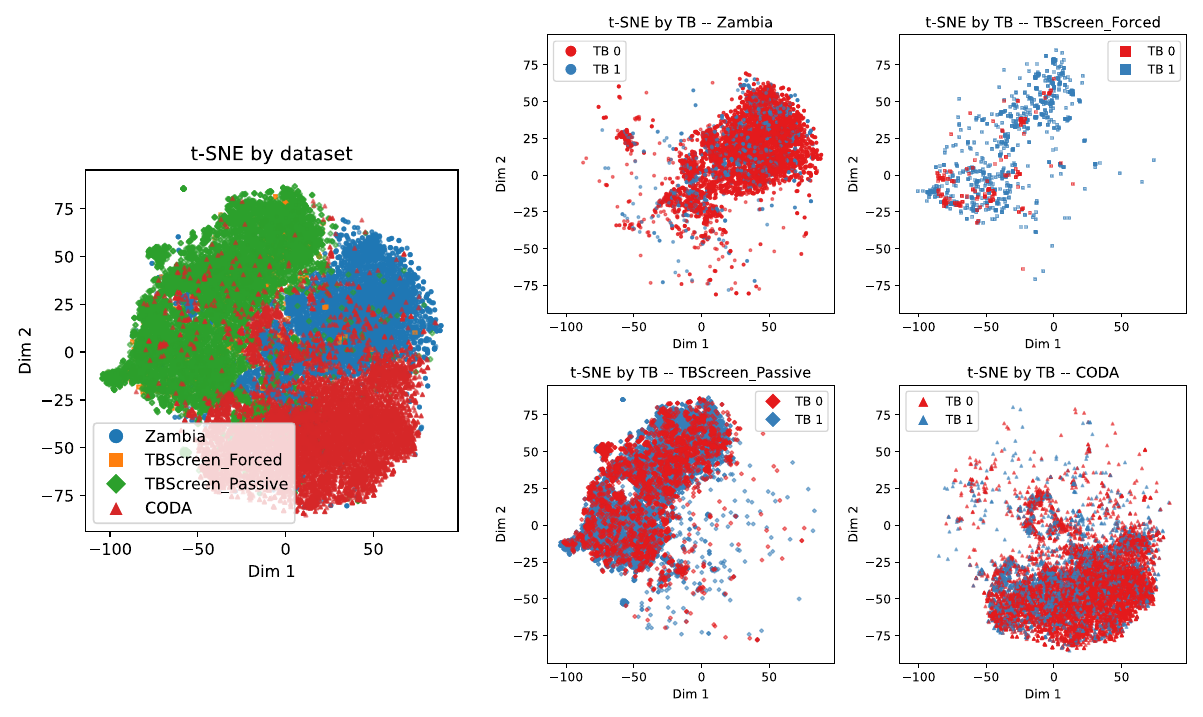}
    \caption{
    \textbf{t-SNE visualization of sample-level feature structure in CLAP embedding by dataset and TB status.}
    Similarly, samples showed strong organization by dataset source, while distribution by TB status seems random.
    }
    \label{fig:tsne_tb_clap}
\end{figure}

\begin{figure}[h]
    \centering
    \includegraphics[width=0.9\linewidth]{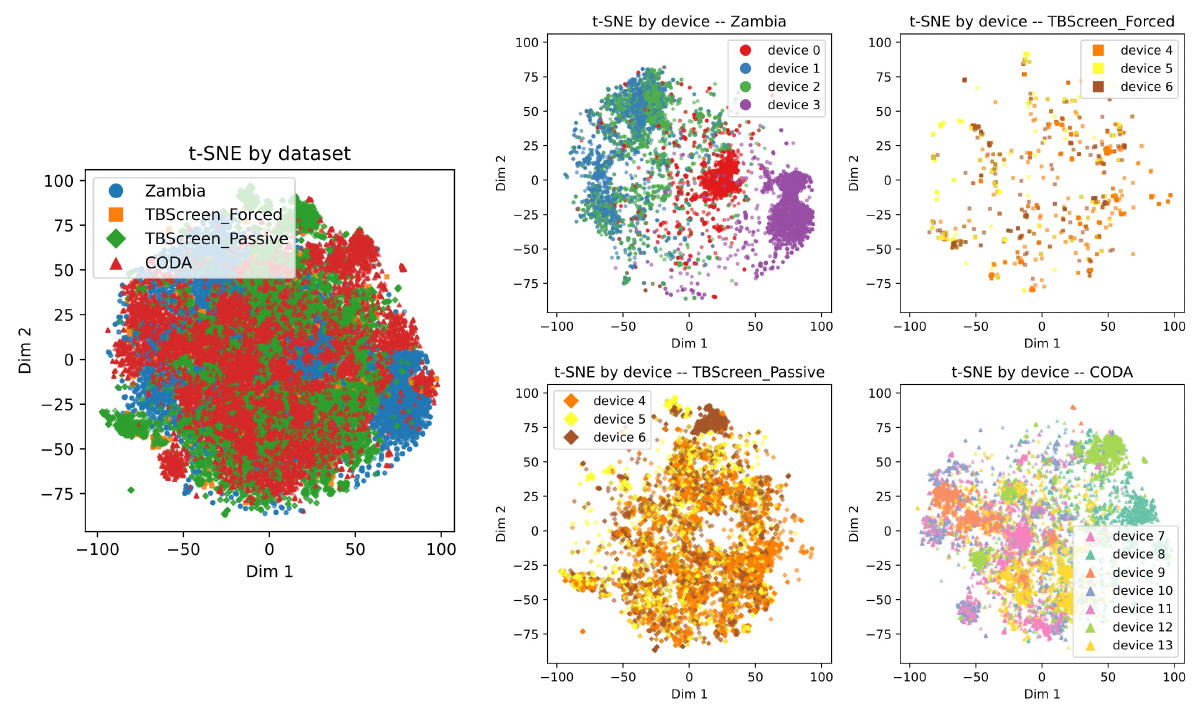}
    \caption{
    \textbf{t-SNE visualization of sample-level feature structure in time/frequency-domain descriptors by dataset and recording device.}
    Similar to CLAP embeddings, samples showed strong organization by dataset source and recording device.
    }
    \label{fig:tsne_device_timefreq}
\end{figure}

\begin{figure}[h]
    \centering
    \includegraphics[width=0.9\linewidth]{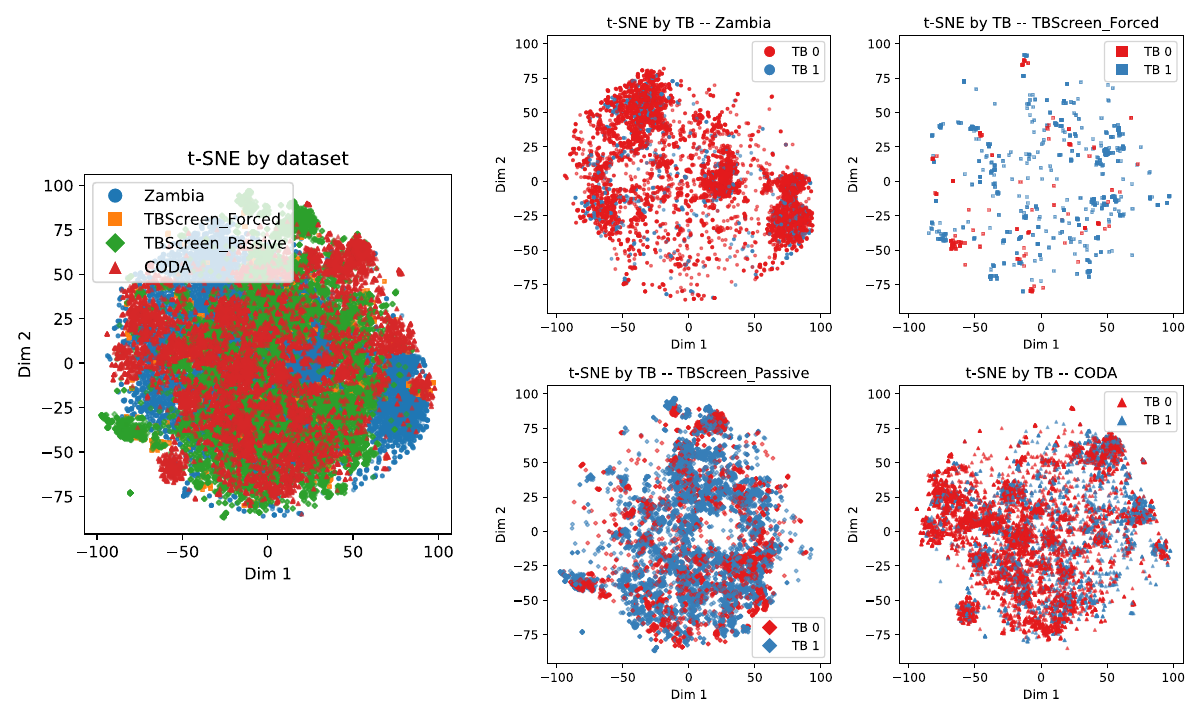}
    \caption{
    \textbf{t-SNE visualization of sample-level feature structure in time/frequency-domain descriptors by dataset and TB status.}
    Similar to previous plot, organization by dataset is strong and organization by TB status is weak.
    }
    \label{fig:tsne_tb_timefreq}
\end{figure}

Fig.~\ref{fig:mmd_timefreq_subgroups} shows the MMD-based MDS visualization of pairwise subgroup distances computed from time/frequency-domain features, with subgroups defined by recording device, location, and TB label. Fig.~\ref{fig:tsne_device_clap} --~\ref{fig:tsne_tb_timefreq} show t-SNE visualizations of CLAP embeddings and time-frequency features, with samples colored by dataset and recording device or by dataset and TB status. Together, these plots complement the CLAP-based representation analysis presented in the main text and confirm that acquisition-related structure dominated both feature spaces.

\section{Domain generalization methods}\label{app:domain_generalization}

Domain shift is a well-studied problem in ML, including in computer vision, speech processing, acoustic scene classification, and broader audio analysis~\cite{wang_generalizing_2023, hassanpour_zonoozi_survey_2023}. A range of domain-generalization and domain-adaptation strategies have been proposed to encourage models to learn features that are less dependent on device or data set source, including MixStyle~\cite{zhou_mixstyle_2024}, frequency-wise MixStyle~\cite{xiao_dg-sed_2025}, Mixup~\cite{zhang_mixup_2017}, LabelGrad~\cite{goodfellow_explaining_2014}, CrossGrad~\cite{shankar_generalizing_2018}, domain-adversarial neural networks~\cite{ganin_domain-adversarial_2016}, kernel-based approaches such as domain-invariant component analysis~\cite{muandet_domain_2013} and scatter component analysis~\cite{ghifary_scatter_2017}, and deep domain confusion~\cite{tzeng_deep_2014}. Motivated by these findings, we explored these strategies in the current study.

\section{Nested cross-validation scheme}\label{app:nested_cv}

\begin{figure}[h]
    \centering
    \includegraphics[width=0.8\linewidth]{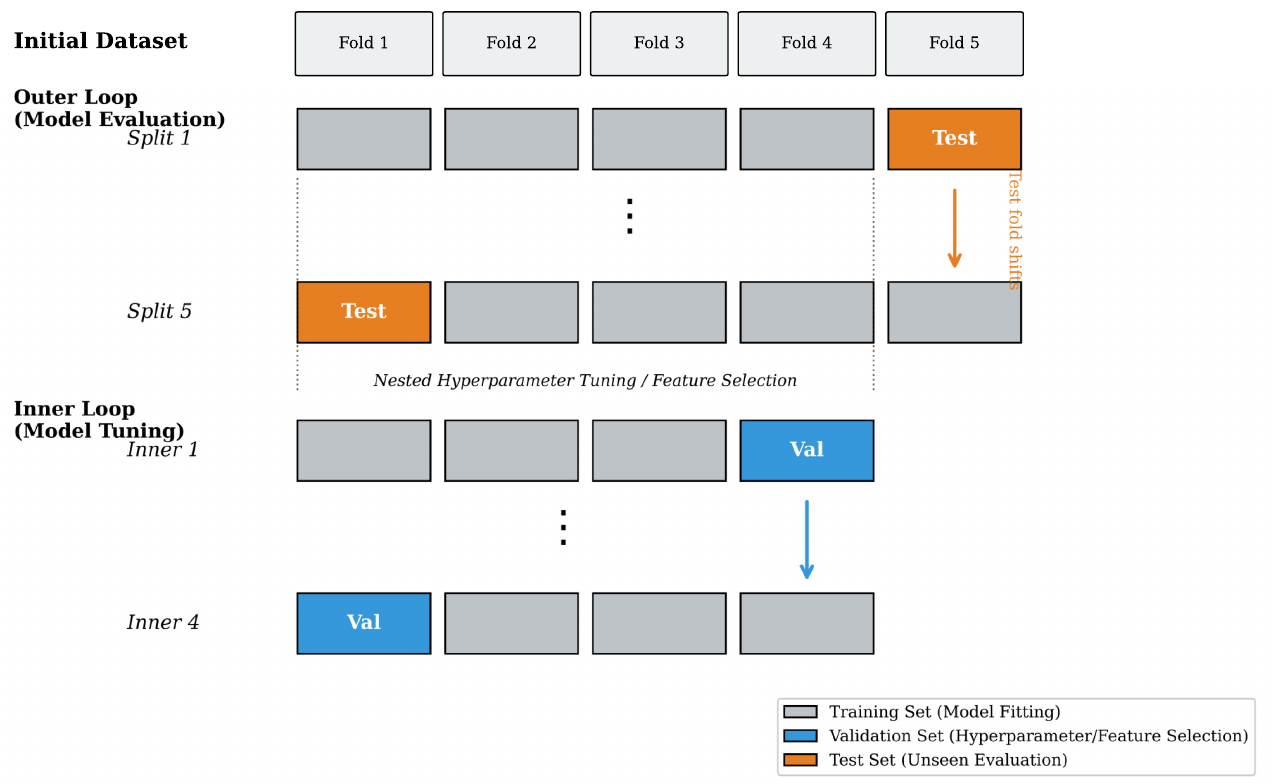}
    \caption{\textbf{Illustration of the nested CV scheme.} In each outer split, one fold is held out for testing, while the remaining four folds are used for model development. Within the development folds, inner CV is used for hyperparameter selection, feature selection, or early stopping.}
    \label{fig:nested_cv}
\end{figure}

Model development and selection within each source dataset used a nested cross-validation (CV) procedure, illustrated in Fig.~\ref{fig:nested_cv}. The outer loop partitioned the source dataset into folds at the subject level and provided an estimate of within-dataset performance, while the inner loop, applied within each outer training partition, was used for model and hyperparameter selection. The models trained on the inner folds were scored on the held-out outer fold, so that no samples used for training and selection contributed to the corresponding performance estimate. Subject-level splitting was enforced throughout to prevent recordings from the same participant appearing in both training and evaluation. The trained models were also evaluated on the remaining datasets as external validation, which were never used during development or tuning. This design separated model selection from performance estimation within each dataset and kept external validation fully independent.

\section{Dataset summary and comparison}\label{app:datasets}
A summary and comparison of the three datasets used in this study are presented in Table~\ref{tab:dataset_comparison}.

The CODA dataset is a multi-country collection assembled for the CODA TB DREAM Challenge~\cite{huddart_dataset_2024}. It contains 733,756 cough sounds from 2,143 adults evaluated for TB in outpatient clinics across India, Madagascar, the Philippines, South Africa, Tanzania, Uganda, and Vietnam. Participants were adults with at least two weeks of cough and the TB evaluation included microbiological tests such as Xpert MTB/RIF Ultra and culture. Coughs were collected using Android smartphones running the Hyfe research application; however, the specific phone model or set of models differed across countries~\cite{huddart_dataset_2024}.

The TBscreen dataset was collected in Nairobi, Kenya in a controlled recording setting~\cite{sharma_tbscreen_2024}. It includes 149 participants with pulmonary TB and 46 controls with other respiratory illnesses, with approximately 33,000 passive coughs and 1,600 forced coughs. Recordings were obtained using three devices: a smartphone (Pixel), a boundary microphone (codec), and a high-end condenser microphone (Yeti). Audio clips with prominent background noise or non-cough respiratory sounds were discarded during annotation.

The Zambia dataset corresponds to the CIDRZ TB dataset used in the HeAR benchmark~\cite{baur_hear_2024}. It was collected by the Centre for Infectious Disease Research in Zambia  (CIDRZ) at three clinical sites in Lusaka district: Chawama, Chainda-South, and Kanyama. Adults were recruited if they had symptoms suggestive of TB, were close contacts with TB patients, or were newly diagnosed with HIV. Each participant was asked to produce four cough events: three single coughs and one episode consisting of consecutive multiple coughs. Recordings were collected using one professional audio recorder and three smartphone tiers spanning different price points: Phone A (Pixel 3a, low-price-range), Phone B (Galaxy A12, mid-price-range), and Phone C (Galaxy A22, high-price-range). Chest X-ray data and associated annotations were collected alongside cough recordings.

\begin{sidewaystable*}[h]
\centering
\caption{Comparison of the three TB cough datasets considered in this study.}
\label{tab:dataset_comparison}
\renewcommand{\arraystretch}{1.2}
\setlength{\tabcolsep}{4pt}
\begin{tabular}{p{1.9cm} p{2.1cm} p{1.9cm} p{1.4cm} p{1.6cm} p{2.9cm} p{3.5cm} p{6.5cm}}
\hline
Dataset & Location & Recording device(s) & Subjects & Cough type & Inclusion criteria & Acquisition & Reference standard / diagnostic work-up \\
\hline

CODA &
India, Madagascar, Philippines, South Africa, Tanzania, Uganda, Vietnam &
Android smartphones &
2,143 adults &
Forced coughs &
Adults with at least two weeks of new or worsening cough, enrolled at outpatient facilities for TB evaluation. &
Participants were asked to cough five times, with each recording lasting 5~s. &
Primary TB status was defined using a microbiologic reference standard based on sputum Xpert MTB/RIF Ultra PCR and mycobacterial culture (Lowenstein--Jensen or MGIT); a sputum Xpert-only reference standard was also provided. \\

TBscreen &
Nairobi, Kenya &
Smartphone (Pixel), boundary microphone (codec), condenser microphone (Yeti) &
195 adults &
Passive and forced cough &
Pulmonary TB cases and controls with other respiratory illnesses recruited in a controlled setting. &
Participants sat in a quiet room for 2~h to collect passive coughs. A subset of participants was additionally asked to provide forced cough recordings. &
Pulmonary TB cases were defined by spontaneous sputum positive on GeneXpert (MTB/RIF or Ultra) and subsequently confirmed by acid-fast bacilli culture; non-TB controls were GeneXpert-negative, had chest radiographs not typical for TB, and were clinically judged to have a non-TB respiratory condition. \\

CIDRZ Zambia / HeAR &
Chawama, Chainda-South, and Kanyama, Zambia &
Pixel 3a, Galaxy A12, Galaxy A22, and audio recorder &
599 adults &
Forced coughs &
Adults with TB symptoms, close contacts of TB patients, or newly diagnosed with HIV &
Participants were asked to generate four cough events: three single coughs and one sequence of consecutive multiple coughs &
TB labels were linked to chest X-ray data and microbiologic evaluation from the parent active case-finding study; in the related Zambia study, the composite reference standard was Xpert MTB/RIF or MTB culture. \\

\hline
\end{tabular}
\end{sidewaystable*}



\section{Time/frequency-domain acoustic features}\label{app:handcrafted_features}

The handcrafted feature set consisted of conventional acoustic descriptors extracted from short-time frames in the time and frequency domains. Features were computed using a window size of 0.02~s and a step size of 0.01~s. For each frame-level feature, the mean and standard deviation across all frames in a cough recording were calculated to obtain a fixed-length representation.

The extracted features included zero-crossing rate (ZCR), frame energy, energy entropy, root mean energy (RME), frame standard deviation, spectral kurtosis, spectral centroid, spectral entropy, spectral flux, spectral rolloff, spectral crest factor, spectral decrease, spectral flatness, spectral slope, spectral skewness, mean frequency, Welch power spectral density statistics (mean and maximum), logarithmic band power, Mel-frequency cepstral coefficients (MFCCs), harmonic ratio (HR), fundamental frequency ($f_0$), and chroma vectors. 

\section{DL backbones}\label{app:deep_backbones}

The DL pipeline evaluated several pretrained backbone models commonly used in audio representation learning.

\textbf{ResNet-18 and ResNet-34.} ResNet-18 and ResNet-34 are convolutional neural network architectures based on residual learning, originally developed for image recognition~\cite{he_deep_2015}. In audio applications, they are commonly used on spectrogram inputs, treating time-frequency representations as images. Their residual connections facilitate optimization and enable moderately deep architectures to learn robust discriminative features.

\textbf{VGGish.} VGGish is a convolutional neural network derived from the VGG architecture and adapted by Google for general-purpose audio classification~\cite{hershey_cnn_2017}. It is pretrained on large-scale audio data and takes log mel-spectrogram patches as input, producing compact embeddings that have been widely used for downstream audio tasks.

\textbf{OPERA.} OPERA is a pretrained foundation model, developed to support health-related audio analysis across multiple tasks and domains~\cite{zhang_towards_2024}. It is designed to learn transferable acoustic representations from large and diverse biomedical and human sound datasets, making it relevant for cough-based screening applications.

\textbf{CLAP.} CLAP is a multimodal representation-learning model trained to align audio signals and text descriptions in a shared embedding space~\cite{elizalde_clap_2023}. Although originally developed for audio-language tasks, its pretrained audio encoder can also be used independently as a general-purpose acoustic feature extractor.

A task-specific classification head was appended to each backbone to predict TB status.

\end{appendices}

\clearpage
\bibliography{sn-bibliography}
